
\def\IR{{\hbox{{\rm I}\kern-.2em\hbox{\rm R}}}}
\def\IB{{\hbox{{\rm I}\kern-.2em\hbox{\rm B}}}}
\def\IN{{\hbox{{\rm I}\kern-.2em\hbox{\rm N}}}}
\def\IC{{\ \hbox{{\rm I}\kern-.6em\hbox{\bf C}}}}

\def\IZ{{\hbox{{\rm Z}\kern-.4em\hbox{\rm Z}}}}
\def\to{\rightarrow}
\def\d{{\rm d}}
\def\underarrow#1{\vbox{\ialign{##\crcr$\hfil\displaystyle
{#1}\hfil$\crcr\noalign{\kern1pt
\nointerlineskip}$\longrightarrow$\crcr}}}
%
\def\d{{\rm d}}
\def\ltorder{\mathrel{\raise.3ex\hbox{$<$}\mkern-14mu
             \lower0.6ex\hbox{$\sim$}}}
\def\lesssim{\mathrel{\raise.3ex\hbox{$<$}\mkern-14mu
             \lower0.6ex\hbox{$\sim$}}}


\def\overlrarrow#1{\vbox{\ialign{##\crcr
$\leftrightarrow$\crcr\noalign{\kern-1pt\nointerlineskip}
$\hfil\displaystyle{#1}\hfil$\crcr}}}

\input phyzzx
\overfullrule=0pt
\tolerance=5000
\overfullrule=0pt
\twelvepoint

\def\bar{\overline}
\twelvepoint
\pubnum{IASSNS-HEP-94/5}
\date{February, 1994}
\titlepage
\title{SUPERSYMMETRIC YANG-MILLS THEORY \break ON A FOUR-MANIFOLD}
\vglue-.25in
\author{Edward Witten
\foot{Research supported in part by NSF Grant
PHY92-45317.}}
\medskip
\address{School of Natural Sciences
\break Institute for Advanced Study
\break Olden Lane
\break Princeton, NJ 08540}
\bigskip
\abstract{By exploiting standard facts about $N=1$ and $N=2$ supersymmetric
Yang-Mills theory, the Donaldson invariants of four-manifolds that admit
a Kahler metric can be computed.  The results are in agreement
with available mathematical computations, and provide a powerful check
on the standard claims about supersymmetric Yang-Mills theory.}
\chapter{Introduction}

\REF\russians{V. Novikov, M. Shifman, A. Vainshtein, and V. Zakharov,
Nucl. Phys. {\bf B229} (1983) 381, 407.}
\REF\cern{D. Amati, K. Konishi, Y. Meurice, G. C. Rossi, and G. Veneziano,
``Non-Perturbative Aspects in Supersymmetric Gauge Theories,'' Phys.
Rep. {\bf 162} (1988) 169.}
\REF\ads{I. Affleck, M. Dine, and N. Seiberg, Nucl. Phys. {\bf B241} (1984)
493.}
In four-dimensional
supersymmetric Yang-Mills theory formulated on flat ${\bf R}^4$,
certain correlation functions are independent of spatial separation
and are hence effectively computable by going to short distances.
This is the basis for one of the most fruitful techniques for studying
dynamics of those theories [\russians,\cern], and gains even more
power when combined with analysis
of instanton corrections to superpotentials [\ads].

\REF\witten{E. Witten, ``Topological Quantum Field Theory,'' Comm. Math.
Phys. {\bf 117} (1988) 353.}
\REF\baulieu{L. Baulieu and I. M. Singer, ``Topological Yang-Mills
Symmetry,'' Nucl. Phys. (Proc. Suppl.) {\bf 15B} (1988) 12.}
\REF\donaldson{S. Donaldson, ``Polynomial Invariants For Smooth
Four-Manifolds,'' Topology {\bf 29} (1990) 257.}
\REF\kd{S. Donaldson and P. Kronheimer, ``The Geometry Of Four-Manifolds,''
Oxford University Press (1990).}
The facts used to establish such results on flat ${\bf R}^4$ also
have global manifestations, especially in the case of $N=2$ and $N=4$
super Yang-Mills theory.  For example, when $N=2$ super Yang-Mills
theory is formulated (in a suitable twisted
fashion which we will recall) on an arbitrary four-manifold $M$,
one can define certain correlation
functions which are independent of the metric of $M$ [\witten].
These correlation
functions are thus, crudely speaking, topological invariants
of $M$.  Actually, since the Yang-Mills action contains derivatives,
its formulation requires a differentiable structure on $M$.  Hence,
the special correlation functions are invariants of $M$ as a differentiable
manifold.  In fact [\witten,\baulieu], they are equivalent to the celebrated
Donaldson invariants [\donaldson,\kd], which are the basis of
much of what is known about smooth four-manifolds.

\REF\friedman{R. Friedman and J. W. Morgan, ``Algebraic Surfaces And
Four-Manifolds: Some Conjectures and Speculations,'' Bull. Am. Math. Soc.
{\bf 18} (1988) 1, {\it Smooth Four-Manifolds And Complex
Surfaces}, to appear.}
\REF\donrev{S. Donaldson, ``Gauge Theory And Four-Manifold Topology,''
preprint.}
\REF\kthree{K. G. O'Grady, ``Donaldson's Polynomials For $K3$ Surfaces,''
J. of Diff. Geom. {\bf 35} (1992) 415.}
\REF\kronheimer{P. Kronheimer and T. Mrowka, ``Recurrence Relations
And Asymptotics For Four-Manifold Invariants,'' submitted to Bull. Amer.
Math. Soc.}
The realization of the Donaldson invariants as correlation functions
of a physical theory leads one to wonder whether physical methods
could be used to calculate them, or conversely, whether
topological computations can be used to constrain the behavior of
physical models.  In this paper, I will present
some results in this direction.  To be specific, I will show how
standard properties of $N=1$ supersymmetric Yang-Mills theory
can be used to determine the Donaldson invariants of those
four manifolds that admit a Kahler metric.  This is an area that
has been much investigated (for reviews see [\friedman,\donrev]),
and the formulas that we will obtain
agree with the known mathematical formulas.  In fact, comparison with
some of those formulas -- especially the formulas of O'Grady
[\kthree] for K3 surfaces
and of Kronheimer and Mrowka [\kronheimer] for so-called four-manifolds of
simple type -- helped considerably in working out the argument
presented in this paper.

\REF\grassmannian{E. Witten, ``The Verlinde Algebra And The Cohomology
Of The Grassmannian,'' IAS preprint HEP-93/41.}
As in a recent study of the Verlinde algebra and the cohomology of
the Grassmannian [\grassmannian], the basic idea will be to exploit
the way a topological theory arises as part of a physical theory.
The topological
correlation functions of interest will arise as specific correlation
functions in a model (essentially $N=1$ supersymmetric Yang-Mills
theory) which has asymptotic freedom, chiral symmetry breaking,
and a dynamically generated
mass gap; the mass gap in particular severely constrains the topological
correlation functions.

To make the subject as accessible as possible, the paper will
be organized as follows.  The problem and its solution are described
in a
relatively non-technical way in \S2.  Many technical details are deferred
for \S3.

\REF\park{J.-S. Park,
``Holomorphic Yang-Mills Theory On
Compact Kahler Manifolds,'' ``$N=2$ Topological Yang-Mills Theory
On Compact Kahler Surfaces,''  ESANAT and Yonsei preprints (1993).}
\REF\ugly{E. Witten, ``Two Dimensional Gauge Theories Revisited,''
J. Geom. Phys. {\bf 9} (1992) 303.}
\REF\johansen{A. Johansen, ``Twisting Of $N=1$ SUSY Gauge Theories
and Heterotic Topological Models,'' Fermilab preprint.}
There have been a couple of other recent papers considering
supersymmetric Yang-Mills theory on Kahler manifolds.  Park
[\park] described many of the relevant general features
and also adapted the two-dimensional construction of [\ugly]
to four dimensions -- where unfortunately this approach does not immediately
give much computational power. Johansen [\johansen] formally wrote
down the twisting of $N=1$ models on Kahler manifolds.

\chapter{Outline Of The Argument}
\REF\wess{J. Wess and J. Bagger, {\it Supersymmetry And Supergravity}
(Princeton University Press, 1983).}

First let us recall some of the basic properties of $N=1$ and $N=2$
super Yang-Mills theory in four dimensions, with gauge group $G$.
In the minimal $N=1$ theory, without extra matter multiplets,
the fields are a gauge field $A_m$ and a gluino field $\lambda$
(a Majorana spinor with values in the adjoint representation).
The Lagrangian is
$$L ={1\over e^2}\int d^4x\sqrt g\left(-{1\over 4}F_{mn}F^{mn}
-\bar\lambda^{\dot\alpha}\sigma^{m}_{\alpha\dot\alpha}
 i D_m \lambda^\alpha\right). \eqn\hurdo$$
Our conventions are as in Wess and Bagger [\wess] and are explained
at the beginning of \S3.
Classically there is a $U(1)$ $R$ symmetry
$$\eqalign{\lambda_\alpha &\to e^{i\gamma}\lambda_\alpha \cr
\bar\lambda_{\dot\alpha}&\to e^{-i\gamma}\bar\lambda_{\dot\alpha}. \cr}
\eqn\urdo$$
Instantons reduce the symmetry group to a subgroup ${\bf Z}_{2h}$
of $U(1)$, where $h$ is the dual Coxeter number of $G$; for $G=SU(2)$,
the main case that we will consider for illustration, the symmetry
group is ${\bf Z}_4$.

This theory is asymptotically free, and its key properties are
believed to be those of conventional QCD: confinement, a mass gap,
and spontaneous chiral symmetry breaking.  The chiral symmetry in question
is the finite group ${\bf Z}_{2h}$ just described.  The standard
conjecture is that this symmetry group is broken down to a ${\bf Z}_2$
subgroup, whose unique non-trivial element is the transformation
$$\lambda_{\alpha}\to -\lambda_\alpha,~~\lambda_{\dot\alpha}\to
-\lambda_{\dot\alpha}.  \eqn\durko$$
This symmetry breaking produces an $h$-fold degeneracy of the vacuum,
which is believed to be the only degeneracy.

In fact, the symmetry in \durko\ is equivalent to a $2\pi$ rotation
in space-time, and so cannot be spontaneously broken if rotation invariance
is unbroken.  On the other hand, additional unbroken symmetries
would forbid fermion masses.  The standard conjectures stated
in the last paragraph are thus in keeping with the experience of ordinary
QCD: in strongly coupled theories of gauge bosons and fermions only,
the unbroken symmetry group is a maximal subgroup of the global symmetry
group that allows fermion masses.

To this minimal $N=1$ theory, it is possible to add matter fields
in the form of ``chiral superfields.'' Such a superfield is a
bose-fermi pair $\Phi =(\phi,\psi)$, of spin $(0,1/2)$,
and transforming in an arbitrary representation of the gauge group.
In addition to the gauge interactions, the chiral superfields
can have additional interactions governed by a ``superpotential.''
The only special case of the superpotential that will be important
in this paper is the following: chiral superfields $\Phi$
in a real representation
of $G$ can have a mass term
$$ \int d^4x \,d^2\theta \,m\Phi^2+{\rm h.c.} \eqn\hudco$$
compatible with supersymmetry.

An important real representation is the adjoint
representation.  If one adds to the $N=1$ theory a chiral superfield
$\Phi$ in the adjoint representation, then the minimally
coupled gauge theory (with vanishing superpotential) in fact has
$N=2$ supersymmetry.  The Lagrangian is
written at the beginning of \S3.  It is possible to
add a bare mass term as in \hudco\ for the $\Phi$ multiplet,
breaking $N=2$ supersymmetry down to $N=1$ supersymmetry.

Though the $N=2$ theory is still asymptotically free, its behavior
is quite different from that of the minimal $N=1$ theory.  The
chiral symmetry group is now, at the classical level, a group
$U(2)$ acting on the pair $(\lambda,\psi)$.  Instantons break
this to a subgroup whose connected component is isomorphic to $SU(2)$;
we will call this internal $SU(2)$ symmetry group $SU(2)_I$.

\REF\seiberg{N. Seiberg, ``Supersymmetry And Non-perturbative Beta
Functions,'' Nucl. Phys. {\bf 208B} (1988) 75.}
In contrast to what one might guess from the behavior of strongly
coupled gauge theories without scalars, it is believed that $SU(2)_I$
is not spontaneously broken and that there is no mass gap.  In fact,
this follows from the form of the classical potential
$$V=\Tr\left[\phi,\bar\phi\,\right]^2, \eqn\fucco$$
with $\phi$ a complex scalar in the adjoint representation.
This potential has ``flat directions,'' with $\langle\phi\rangle\not=0$,
breaking the gauge symmetry down to $U(1)^r$ ($r$ being the rank of $G$),
and leaving a weakly  coupled abelian theory, in which all charged fields
are massive.  This behavior holds in the quantum theory at least
in the weakly coupled region of large $\langle\phi\rangle$,
since $N=2$ supersymmetry
permits no perturbations that would lift the vacuum degeneracy in the
flat directions.  The instanton contributions in the weakly coupled
theory for large
$\langle\phi\rangle$ have been analyzed by Seiberg [\seiberg].

The goal in what follows is to use facts about the topology of four
dimensions to retrieve the conclusion just stated that the minimal
$N=2$ theory does not have a mass gap, and conversely
to use the mass gap and related properties of the $N=1$ theory
to deduce conclusions about the topology of four dimensions.

\section{Twisting}

First we will recall how one constructs from the $N=2$ super Yang-Mills
theory a twisted topological field theory.  For more details,
the reader can consult [\witten] and also \S3 below.

The rotation group $K$ in four dimensional Euclidean space
is locally $SU(2)_L\times SU(2)_R$.
In addition, the connected component of the global symmetry group
of the $N=2$ theory is, as explained above, $SU(2)_I$.  The theory,
when formulated on a flat ${\bf R}^4$, therefore has a global symmetry
group
$$H=SU(2)_L\times SU(2)_R\times SU(2)_I. \eqn\jjury$$
The supercharges $Q_\alpha{}^i$ and $Q_{\dot \alpha i}$
transform under $H$ as $(1/2,0,1/2)$
and $(0,1/2,1/2)$, respectively.

Apart from its standard embedding in $H$ (as $SU(2)_L\times SU(2)_R$),
$K$ also has nonstandard embeddings.  Let $SU(2)_R{}'$ be a diagonal
subgroup of $SU(2)_R\times SU(2)_I$, and let
$$K'=SU(2)_L\times SU(2)_R{}'.\eqn\andlet$$
Then $K'$ is isomorphic to $K$, and we can think of the $N=2$ super
Yang-Mills theory as a Poincar\'e invariant theory with rotations acting
by $K'$.  Use of $K'$ instead of $K$ to generate rotations is natural
if one replaces the standard stress tensor $T$ of the theory by a modified
stress tensor $T'$ (which differs from $T$
by addition of a derivative term that does not contribute to the translation
generators).

But this substitution changes
the physical interpretation of the theory considerably.  All
fields,
commuting or anticommuting, have integer spin with respect to $K'$.
In particular, this is so for the supercharges, which under
$SU(2)_L\times SU(2)_R{}'$ transform as $(1/2,1/2)\oplus (0,1)\oplus
(0,0)$.

The $(0,0)$ component of the supercharge is the rotation
invariant object $Q=\epsilon^{\dot\alpha i}Q_{\dot\alpha i}$.
By virtue of the supersymmetry algebra, $Q$ obeys $Q^2=0$; this
follows from the fact that $Q^2$ transforms as $(0,0)$ while the only
bosonic operators in the supersymmetry algebra are the momenta,
transforming as $(1/2,1/2)$.
Moreover, it is possible to find a component $S$ of the underlying
supercurrent such that the modified stress tensor $T'$ can be written
$$T'{}_{\mu\nu}=\{Q,S_{\mu\nu}\} .\eqn\imdobo$$
These facts ensure that with $Q$ regarded as a BRST operator, and $T'$
used as the stress tensor when coupling to gravity,
the theory can be interpreted as a topological field theory.

Indeed [\witten], more or less because $Q$ is Lorentz invariant in the $K'$
sense,
the $N=2$ theory can be formulated on an arbitrary four manifold $M$
in such a way as to preserve the existence of the fermionic symmetry
$Q$ and the basic relations $Q^2=0$ and \imdobo.  At this stage the
twisted theory comes into its own. On flat ${\bf R}^4$, the twisted
theory is merely a different way of looking at the usual physical theory;
but as soon as we formulate the theory on a curved space-time, the twisted
theory is really different, because of the use
of a different stress tensor.

\section{Operators}

The important part of the BRST symmetry (in the sense that other
fields do not enter in the construction of observables)  is
$$\eqalign{\delta A_m & = i\epsilon \lambda_m \cr
           \delta \lambda_m & = - \epsilon D_m \phi \cr
           \delta \phi & = 0 .   \cr   }   \eqn\icco$$
Here $A$ is the gauge field, $\lambda_m$ is a linear combination of the
fermions (the details will be described in \S3),
$\phi$ is a scalar, and $\epsilon$ is an anticommuting parameter.
One easily verifies from this formula that, up to a gauge
transformation, $Q^2=0$ in acting on these fields.  (The multiplet
\icco\ has a mathematical interpretation in terms of the equivariant
cohomology of the gauge group acting on the space of connections,
but that need not concern us here.)

\def\co{{\cal O}}
Since $\phi$ is BRST invariant, but cannot be written as $\{Q,\dots\}$,
the obvious BRST invariant observables are of the form $P(\phi)$,
with $P$ an invariant polynomial on the Lie algebra of $G$.
For $G=SU(2)$, the basic invariant polynomial is the quadratic Casimir
operator.  (All others are polynomials in this and lead to nothing
essentially new.)  So the basic $Q$-invariant observable is
$$\co(x)=\co^{(0)}(x) = {1\over 8\pi^2}\Tr \phi^2(x), \eqn\barnaby$$
with $\Tr$ the trace in the fundamental two dimensional representation
of $SU(2)$.  More generally, for a Lie group of rank $r$, there are
$r$ independent Casimir operators and accordingly $r$ independent
operators generalizing $\co^{(0)}(x)$.

So the first example of a topological correlation function is
$$\left\langle \co(x_1)\co(x_2)\dots \co(x_s)\right\rangle,  \eqn\guffrey$$
with arbitrary distinct points $x_i$.  (There is no singularity in
the $\co(x)\co(y)$ operator product for $x\to y$, so one could in fact
let some of the $x_i$ coincide, but it is more convenient to keep them
distinct.)
This correlation function
is independent of the metric of $M$ because of \imdobo; and so
it must be independent of the choice of the $x_i$, because
up to a change in metric on $M$,
there is no invariant information in the choice
of distinct points $x_1\dots x_s\in M$.
To prove more explicitly that \guffrey\ is independent of the $x_i$,
note that although $\co$ cannot be written as $\{Q,\dots\}$,
its derivative can be:
$${\partial\over\partial x^m}\co =\{Q,{\Tr \phi\lambda_m\over 4\pi^2}
 \}.  \eqn\hurry$$

I write this equation as $d\co^{(0)} =\{Q,\co^{(1)}\}$, where
$\co^{(1)}$ is an operator-valued one-form $\co^{(1)}=
\Tr \phi\lambda_m dx^m/4\pi^2$.
It is the beginning of a descent procedure.  One recursively finds
operator-valued $k$-forms $\co^{(k)}$ such that
$$d \co^{(k)}=\{Q,\co^{(k+1)}\}. \eqn\icco$$
For full details on this, see [\witten].  \icco\
means that $\co^{(k)}$ is $Q$-invariant up to an exact form
and is closed up to $\{Q,\dots\}$.  Accordingly, if $\Sigma$
is a $k$-dimensional submanifold of $M$ (or more generally a $k$-dimensional
homology cycle), then
$$I(\Sigma)=\int_\Sigma {\cal O}^{(k)} \eqn\hicco$$
is BRST invariant and depends only on the homology class of $\Sigma$.

In practice Donaldson theory is most intensively studied on simply
connected four manifolds $M$.  In that case, $k$-dimensional homology
cycles $\Sigma$ exist only for $k=0,2,4$.  For $k=0$, $\Sigma$ is a point
$x\in M$, and $I(\Sigma)$ is simply our friend $\co(x)$.  For $k=4$,
$\Sigma$ must be $M$ itself.  For $\co$ constructed from the quadratic
Casimir operator, $I(M)$ turns out to be simply an elementary topological
invariant, the instanton number $I(M)=\int_M\Tr F\wedge F/8\pi^2$.
Therefore, for $G=SU(2)$, interesting quantum operators do not arise
for $k=4$.  Interesting operators with $k=4$ do arise
for other groups, which have higher Casimir operators.

Hence for
$SU(2)$, apart from $k=0$,  we focus on the case that $k=2$ and that
$\Sigma\subset M$ is an oriented two dimensional submanifold of $M$.
For each such $\Sigma$, we get a BRST invariant operator $I(\Sigma)$,
which up to $\{Q,\dots\}$
depends only on the homology class of $\Sigma$.  If we denote
the components of the operator-valued two-form $\co^{(2)}$ as
$$Z_{mn}(x)={1\over 4\pi^2}
\Tr \left(\phi F_{mn}-i\lambda_m\lambda_n\right)\eqn\pondd$$
then
$$I(\Sigma)=\int_\Sigma Z_{mn}d\sigma^{mn}.      \eqn\jorry$$
The correlation functions
$$\left\langle
\co(x_1)\dots \co(x_r)I(\Sigma_1)\dots I(\Sigma_s)\right\rangle \eqn\gorry$$
are the celebrated Donaldson invariants of smooth four-manifolds.

\REF\tdgt{E. Witten, ``On Quantum Gauge Theories In Two Dimensions,''
Commun. Math. Phys. {\bf 141} (1991) 153.}
There is actually a small subtlety here: the Donaldson invariants
as usually defined by topologists are larger by a universal
constant factor than the correlation functions just introduced.
This factor is the number of
elements in the center of the gauge group; we will call the center
$Z(G)$ and let $\#Z(G)$ denote the number of its elements.
Thus if $\langle~~~\rangle_T$ denotes the
Donaldson invariants as usually defined topologically, then
$$
\left\langle
    \co(x_1)\dots \co(x_r)I(\Sigma_1)\dots I(\Sigma_s)\right\rangle_{T}=
\#Z(G)\cdot \left\langle
    \co(x_1)\dots \co(x_r)I(\Sigma_1)\dots I(\Sigma_s)\right\rangle
.\eqn\smallcorrection$$
The reason for this is that to recover the conventional topological
definition of Donaldson's invariants from the path integral
(by a semi-classical evaluation of the path integral, discussed in
[\witten,\baulieu]) one must delete a factor of $1/\#Z(G)$ that
appears in the properly normalized path integral and is explained
in [\tdgt, p. 159].
In fact, in the Fadde'ev-Popov definition of the path integral,
one divides  by the volume of the gauge group, taking account of the
center of the group even though it acts
trivially on the space of connections; but topologists usually omit
to divide by the order of the center.

\subsection{The Dimension Of Moduli Space}

I will not attempt here a systematic account of the field theoretic
approach to Donaldson theory.  However, one point is so fundamental
that it requires mention.

The classical $N=2$ theory has a $U(2)$ symmetry, acting on
the two fermi fields $(\lambda_\alpha,\psi_\alpha)$. So far we
have exploited the anomaly-free $SU(2)$, in twisting the model.
The center $U(1)\subset U(2)$ is anomalous; nevertheless it plays
an essential role.  Let us call this quantum number $U$.

As explained in [\witten], on a given four-manifold and for a given
instanton number, the total violation of $U$, which we will call
$\Delta U$, equals the dimension of the Yang-Mills instanton moduli
space ${\cal M}$.  For $SU(2)$ this is
$$\Delta U=\dim {\cal M}=8k -{3\over 2}\left(\chi+\sigma\right),
\eqn\moddim$$
with $k$ the instanton number and
$\chi$ and $\sigma$ the Euler characteristic and signature of $M$.
(The quantity $(\chi+\sigma)/2$ is always integral.)

On the other hand, the operators $\co$ and $I(\Sigma)$ have
$U=4$ and $U=2$, respectively, so a correlation function
$$\left\langle\co(x_1)\dots\co(x_r)I(\Sigma_1)\dots I(\Sigma_s)\right\rangle
       \eqn\dormo$$
vanishes unless
$$ 4r+2s =\dim{\cal M}= 8k -{3\over 2}\left(\chi+\sigma\right).\eqn\ippso$$

\section{Role Of A Mass Gap}

$N=2$ supersymmetric Yang-Mills theory does not have a mass gap
-- for reasons explained earlier.  Nevertheless, we will ultimately
perform calculations by perturbing to a situation in which there
is a mass gap.  To see how the mass gap enters, we will first
assume -- counter-factually -- that the $N=2$ theory does have a mass
gap, and see what computations become possible.

Actually, some of the arguments (those that involve $I(\Sigma)$
as well as $\co$) require a somewhat stronger assumption.
We will assume that the
ground state structure, on any space-time manifold, is the same as
it is in the bulk
theory.  This means roughly that the vacua can be completely
labeled by gauge-invariant local observables.
The only known theories with mass gap in which
this is not so are theories with unbroken and unconfined
gauge symmetries, in which measurements of Wilson lines or global holonomies
are needed to distinguish the vacua.  Thus the assumption that we will
exploit is essentially
that the $N=2$ theory has a mass gap, and no unbroken, unconfined
gauge symmetries.

We will also temporarily make a further simplification and assume
that there is only one vacuum. This is a minor point and will eventually
be corrected by introducing a sum over vacua.

\subsection{Consequences Of The Assumption}

With our assumptions, let us determine the Donaldson invariants.
First we consider the zero point function -- that is the partition
function in the absence of any operator insertions.
We will use the symbol $\langle X\rangle$ to refer to an
unnormalized path integral on the four-manifold $M$ with an insertion of
the operator $X$, so the partition function will be denoted
as $\langle 1\rangle$.  When we need to refer to the normalized vacuum
expectation value of $X$ in the infinite volume limit (on a flat
${\bf R}^4$), we will write this as $\langle X\rangle_\Omega$, with
$\Omega$ being the infinite volume vacuum state.

We use our liberty to pick a convenient metric on $M$.  We begin
with any metric $g$, and then scale it up by $g\to tg$, with $t$ a positive
constant, and consider what happens for $t\to \infty$.  The metric on
$M$ becomes everywhere nearly flat.  In a theory with a mass gap,
the response to a background gravitational field is given by an effective
action that can be expanded as a sum of local operators.
In fact,  the partition function is
$$\langle 1\rangle =\exp(-L_{eff}), \eqn\gurf$$
where
$L_{eff}$ has an expansion
$$L_{eff}=\int d^4x\sqrt g\left(u+vR +w R^2 +\dots\right), \eqn\urf$$
with constants $u,v,w,\dots$ and local operators $1,R,R^2,\dots$ ($R$ is the
Ricci scalar of $M$).
This is the general structure that would arise in any local theory with
a mass gap, but in the case at hand the expansion is severely restricted
by topological invariance.
The above expansion is valid for large $t$.  If $U$ is an operator
of dimension $n$, then $\int d^4x\sqrt g U$ scales as $t^{4-n}$,
so the terms written explicitly in \urf\ make contributions of
order $t^4, t^2$, and 1.  Topological
invariance means that in the particular case at hand, the expansion
must be independent of $t$, so only operators
of dimension four can appear.
The only topological invariants of a four-manifold
that can be written as the integral
of a local operator are the
Euler characteristic $\chi$ and the signature $\sigma$, so
$$\langle 1\rangle =\exp(a\chi+b\sigma) \eqn\uddu$$
with some universal constants $a$ and $b$.  In the case of $SU(2)$
Donaldson theory, we will eventually determine $a$ and $b$ by comparing
to computations on some particular four-manifolds.

Now we include the local operator $\co$ and consider its correlation
functions
$$\left\langle \co(x_1)\dots\co(x_r)\right\rangle. \eqn\delmo$$
Upon scaling up the metric, we can assume that the $x_i$ are far
apart from one another and inserted in a region of $M$ that is essentially
flat on a scale much larger than any Compton wavelength in the theory.
So via cluster decomposition, $\co$ can be replaced by its vacuum
expectation value $\langle\co \rangle_\Omega$:
$$\langle\co(x_1)\dots\co(x_r)\rangle=\langle\co
\rangle_\Omega{}^r\cdot\langle 1\rangle.\eqn\orm$$
This can conveniently be written
$$\langle \exp(\lambda\co)\rangle
=\langle 1\rangle\cdot \exp(\lambda\langle
\co\rangle_\Omega), \eqn\dorm$$
with $\lambda$ a complex parameter.

The arguments become more
interesting if we consider the operator $I(\Sigma)$.
I will first show that given the
assumptions,
the one point function of this operator vanishes,
$$\langle I(\Sigma)\rangle = 0, \eqn\huxxo$$
for every four manifold $M$ and every two dimensional surface
(or homology cycle) $\Sigma\subset M$.
In fact,
$$\langle I(\Sigma)\rangle =\int_\Sigma d\sigma^{mn}\langle Z_{mn}\rangle
           , \eqn\foff$$
and the desired conclusion will follow from suitable properties
of the one point function
$$ \langle Z_{mn}(x)\rangle.       \eqn\goff$$
$Z_{mn}(x)$
is a well-defined, gauge-invariant local operator of the sort
one usually studies in physical applications of Yang-Mills theory.
Its one point function is a standard sort of Yang-Mills observable
but is not a topological invariant; only integrated
expressions like \foff\ are such invariants.
In studying \goff\ as a prelude to analyzing \foff,
we are taking the main step in this paper: constraining the topological
correlation functions by exploiting the way they
arise as special correlators in a physical theory.

Clearly, on ${\bf R}^4$ with a flat metric, $\langle Z_{mn}\rangle =0$
by Lorentz invariance.  To determine what happens on a general
four-manifold, we again consider an arbitrary metric $g$ scaled up
by $g\to tg$ with $t\to\infty$.  Since
the area of $\Sigma$ grows like $t^2$ for $t\to \infty$,
to show that the integrated expectation value \foff\ vanishes,
it suffices to show that for every $x\in M$, $\langle Z_{mn}(x)\rangle$
vanishes for large $t$ faster than $1/t^2$.  This is so as the mass gap
lets us write an expansion for $\langle Z_{mn}(x)\rangle$
in terms of local invariants of the Riemannian geometry of $M$.
The expansion looks like
$$\langle Z_{mn}\rangle =D_m R D_n D_s D^s R \mp \dots       \eqn\onogo$$
Since all possible terms have dimension considerably greater than
two, the expectation value vanishes faster than $1/t^2$.

Now, we move on to compute the two point function of $I(\Sigma)$.
Let $\Sigma_1,\Sigma_2$
be two two-dimensional submanifolds of $M$.
The desired two point function is
$$\left\langle  I(\Sigma_1)I(\Sigma_2)
       \right\rangle =\int_{\Sigma_1\times \Sigma_2}
\left\langle Z_{mn}(x)Z_{pq}(y)\right\rangle d\sigma^{mn}(x)d\sigma^{pq}(y).
            \eqn\yogo$$
As before, we analyze this by scaling up the metric by $g\to tg$,
with $t\to\infty$.  For $x\not=y$, $\langle Z_{mn}(x)Z_{pq}(y)\rangle$
vanishes faster than $1/t^4$, by the same reasoning as above.  The
only possible surviving contribution would be a contribution that for
$t\to\infty$ is localized where $x=y$, that is at the intersection points
of $\Sigma_1$ and $\Sigma_2$.

By perturbing $\Sigma_1$ and $\Sigma_2$ slightly
within their homology (and even homotopy) classes, we can assume that
they meet transversely at finitely many points $w_1,\dots,w_n$.
We can also assume that the metric of $M$ is flat in a neighborhood of
the $w_i$ and that the $\Sigma_a$ look locally like intersections of
coordinate hyperplanes.  The contribution to \yogo\ which for $t\to\infty$
is localized at one of the intersection points $w_i$ can only
depend on local invariants of the behavior of the $\Sigma_a$ near $w_i$.
The only such invariant is the relative orientation with which $\Sigma_1$
and $\Sigma_2$ meet at $w_i$.
On symmetry grounds, the local contribution in \yogo\ is proportional
to this relative orientation. (The operator $I(\Sigma)$ is defined
by integrating a differential form over $\Sigma$ and so changes sign
if the orientation of $\Sigma$ is reversed; reversing the orientation
of one of the $\Sigma_a$ changes the relative orientation with which
$\Sigma_1$ and $\Sigma_2$ meet at $w_i$.)  So -- given our assumption --
the sum of the local
contributions in \yogo\ is proportional to the algebraic intersection number
of the $\Sigma$'s:
$$\left\langle  I(\Sigma_1)I(\Sigma_2)\right\rangle
=\eta\cdot\#(\Sigma_1\cap\Sigma_2)
\cdot \langle 1\rangle.       \eqn\ombbo$$
Here $\eta$ is an unknown universal constant.  This formula is valid
even for $\Sigma_2=\Sigma_1$; one sees this by perturbing $\Sigma_2$
within its homotopy class to a new surface $\Sigma_1{}'$ with only
transverse intersections with $\Sigma_1$.

By a generalization of this reasoning, if $\Sigma_1\dots \Sigma_s$ are
independent two-dimensional submanifolds of $M$, then the mass gap implies
$$\left\langle \exp\left(\sum_a \alpha_a I(\Sigma_a)\right) \right\rangle
 = \exp\left({\eta\over 2}\sum_{a,b}\alpha_a\alpha_b\#(\Sigma_a\cap \Sigma_b)
\right)\cdot \langle 1\rangle. \eqn\cvoovv$$
Indeed, the $\Sigma_a$ can be assumed to have only pairwise intersections;
the right hand side of \cvoovv\ comes from the local contributions
of the intersection points.  The $I(\Sigma_a)$ are analogous to
quantum fields with Gaussian correlations controlled by the two point
function \ombbo.

No new phenomena occur if we consider correlators with $\co$ and $I(\Sigma)$
together, since the points at which $\co$ is inserted can be taken
disjoint from the surfaces $\Sigma_a$ -- and then cluster decomposition
can be used, as before, to replace the $\co$'s by their vacuum expectation
values.  So we can combine and \dorm\ and \cvoovv\
to get
$$\left\langle \exp\left(\sum_a\alpha_aI(\Sigma_a)+\lambda\co\right)
\right\rangle
 = \exp\left({\eta\over 2}\sum_{a,b}\alpha_a
\alpha_b\#(\Sigma_a\cap \Sigma_b)
+\lambda\langle \co\rangle\right)\cdot \langle 1\rangle. \eqn\cvor$$

\subsection{Several Vacua}

It remains to generalize this formula to take account of the
possibility that the theory has not one vacuum state $\Omega$
but several.  Assume existence of a discrete
set of vacua $\Omega_\rho,$ with $\rho$ ranging over some finite
set $S$.  To generalize \cvor, we simply write a sum over
contributions of the $\Omega_\rho$, allowing for the fact that
$\eta$, $\langle\co\rangle $, and the constants $a$ and $b$ of
equation \uddu\ may depend on the choice of vacuum.
We get
$$\left\langle \exp\left(\sum_a\alpha_aI(\Sigma_a)+\lambda\co\right)
\right\rangle  =\sum_{\rho\in S}
 C_\rho\exp\left({\eta_\rho
\over 2}\sum_{a,b}\alpha_a\alpha_b\#(\Sigma_a\cap \Sigma_b)
+\lambda\langle \co\rangle_\rho\right). \eqn\cvovvo$$
Here I have written the partition function $\langle 1\rangle_\rho$
in the $\rho$ vacuum as $C_\rho$; by \uddu\ it is of the
form
$$C_\rho=\exp\left(a_\rho\chi +b_\rho\sigma\right) \eqn\unicon$$
with some universal constant $a_\rho$ and $b_\rho$.

\section{Restriction To Kahler Manifolds And $N=1$}

Upon comparison to mathematical computations of Donaldson invariants
in special cases, one finds that
these formulas are in fact false.  Therefore, the $N=2$ theory does
not obey our assumptions; this comes as no surprise, in view of the flat
directions in the classical potential which ensure that the $N=2$
theory does not have a mass gap.  Were \cvovvo\ valid, Donaldson theory
would detect only familiar, classical invariants of four-manifolds
and so would not have its usual mathematical applications.

However, in certain respects, the formulas are nearly true at least
for $G=SU(2)$ where mathematical computations have been performed.
For instance,
they are true (with minor modification) for $K3$ surfaces
according to results of [\kthree].  For a large class of four manifolds
(those of ``simple type''), \cvovvo\ gives the
right asymptotic  behavior for
large values of $\alpha_a$, according to results of Kronheimer and
Mrowka [\kronheimer].  (To be more precise, according to [\kronheimer],
the Gaussian function of the $\alpha_a$ that appears on the right
hand side must be multiplied by a sum of
exponentials, which are ``subleading'' but contain all of the subtle
information in the theory.)  Why does our simple computation
based on a faulty assumption enjoy these successes?

\REF\gompf{R. Gompf and T. Mrowka, ``A Family Of Non-complex Homotopy
$K3$ Surfaces,'' to appear in Annals of Math.}
The thesis of the present paper will be that this is so because
the assumptions {\it are} valid for $N=1$ super Yang-Mills theory --
and this is almost as good.  To be more precise, in what follows
we will show how $N=1$ super Yang-Mills theory can be used to
study Donaldson theory in
the case that the four-manifold $M$ is a Kahler manifold
(that is, a manifold admitting a Kahler metric).
This is in fact a very basic case in Donaldson theory and
the main arena of actual computations.
For some time it was conjectured that the Donaldson invariants are
zero except for four-manifolds admitting a Kahler metric; this is now known
to be false [\gompf], but the known counter-examples are still closely
related to Kahler manifolds.

\subsection{The Twisted Theory On A Kahler Manifold}

For our purposes, the definition of a four-dimensional
Kahler manifold is that
it is a Riemannian
four-manifold on which the holonomy group is not the usual
$SU(2)_L\times SU(2)_R$, but is contained in $SU(2)_L\times U(1)_R$,
with $U(1)_R$ being a subgroup of $SU(2)_R$ isomorphic to $U(1)$.

The two dimensional representation of $SU(2)$ has a single anti-symmetric
invariant; given two doublets $v_\alpha,$ $w_\beta$ this invariant
can be written as $\epsilon^{\alpha\beta}v_\alpha w_\beta$.  Upon
restriction to  a $U(1)$ subgroup, the same representation also admits
a symmetric invariant, which we might write as $\delta^{\alpha\beta}
v_\alpha w_\beta$.  The implication of this for the twisted $N=2$
model is as follows.  While on a general four-manifold the twisted
theory has a single
BRST operator $Q=\epsilon^{\dot\alpha i}Q_{\dot \alpha i}$,
on a Kahler manifold there is also a second one $Q'=\delta^{\dot\alpha i}
Q_{\dot\alpha i}$.

One way to formulate this is that on a Kahler manifold, one can write
$$ Q= Q_1+Q_2, \eqn\hoffer$$
where $Q_1$ is the part of $Q$ inside one $N=1$ subalgebra and $Q_2$
is the part in the second.
Exactly how the Kahler reduction in structure group makes possible
the decomposition in \hoffer\ will be clarified in \S3.
These operators obey
$$Q_1{}^2=Q_2{}^2=\{Q_1,Q_2\}=0 , \eqn\occo$$
and the observables of Donaldson theory, such as $\co$ and $I(\Sigma)$,
are annihilated by both $Q_1$ and $Q_2$.

In explicit computations via instantons, $Q$, $Q_1$, and $Q_2$
correspond to the de Rham exterior derivative $d$ and to the
$\bar\partial$ and $\partial$ operators of instanton moduli space.
It is a familiar fact in cohomology of smooth, compact, Kahler manifolds
that the (complex-valued) de Rham cohomology is isomorphic to the
$\bar\partial$ cohomology.  Somewhat analogously, in the Donaldson
theory of Kahler manifolds, one can disregard the $Q_2$ symmetry
and keep track of the topological behavior using $Q_1$ alone; this will
be explained in \S3.1.

Since $Q_1$ lies in an $N=1$ subalgebra,
perturbations that break $N=2$ supersymmetry
down to $N=1$ can preserve $Q_1$ invariance.
Is there such a perturbation that will simplify the problem?

\section{The Mass}

This question is no sooner asked than answered.
Temporarily we work on a flat ${\bf R}^4$, with Euclidean
coordinates $y^1,\dots , y^4$.  Of course, we pick a complex
structure, determined say by complex coordinates $z_1=y^1+iy^2$, $z_2
=y^3+iy^4$, to regard ${\bf R}^4$ as a Kahler manifold.

As was explained
at the outset of this section, the $N=2$ theory, regarded as an $N=1$
theory, has the following multiplets: there is a gauge multiplet
$(A,\lambda)$ and a chiral supermultiplet $\Phi=(\phi,\psi)$ in
the adjoint representation.  The chiral supermultiplet can have
a bare mass
$$\Delta L = -m\int d^4x d^2\theta \Tr\Phi^2-{\rm h.c.} \eqn\immmo$$
preserving $N=1$ supersymmetry.
This can be written in the general form
\foot{That is, it can be written this way in the case
of a global Kahler manifold, as we will see
in \S3. On flat
${\bf R}^4$, the term $\sum_a\alpha_aI(\Sigma_a)$ should better be written
$\int_{{\bf R}^4}\omega\wedge Z$, where $\omega=dz_1\wedge dz_2$.  Globally,
the two formulations are equivalent up to $\{Q_1,\dots\}$.}
$$\sum_a\alpha_aI(\Sigma_a)
+\{Q_1,\dots\}, \eqn\huber$$
 so up to $\{Q_1,\dots\}$, by adding the mass term we are just
shifting the Lagrangian by a linear combination of the observables that
we want to study anyway.

The $N=2$ theory perturbed by the $N=1$ invariant bare mass term
\immmo\ has a mass gap for the following reasons.  The perturbation
certainly gives a mass to the $\Phi$ multiplet, so we are reduced
at low energies to a minimal $N=1$ system.  That system in turn
has a dynamically generated
mass gap, at a much lower scale  if $m$ is large.

What are the global symmetries of the perturbed system? The minimal
$N=1$ model has a ${\bf Z}_{2h}$ symmetry described earlier.  The
only symmetry of the $\Phi$ multiplet is $\Phi\to -\Phi$, which generates
a group that we will call ${{\bf Z}_2}'$.  The
overall global symmetry group is hence ${\bf Z}_{2h}\times {{\bf Z}_2}'$.
The pattern of symmetry breaking is easy to work out.
${\bf Z}_{2h}$ is broken to ${\bf Z}_2$, as explained earlier, while
the symmetry ${{\bf Z}_2}'$ that acts only on fields with a positive
bare mass (which can be taken large) is unbroken.  The unbroken
symmetry group is therefore ${\bf Z}_2\times {{\bf Z}_2}'$.

\subsection{The Mass Term On A Curved Kahler Manifold}

If we consider not flat ${\bf R}^4$ but a general Kahler manifold $M$,
an important subtlety arises in the description of the mass term.
The subtlety arises because of the twisting by $U(1)_R$.  The
mass term in
\immmo\ is defined using a measure $d^4x\,d^2\theta$; the factor
$d^2\theta$ carries $U(1)_R$ charge (in the twisted model)
and there is no natural
way to fix the space-time
dependence of its phase on a curved Kahler manifold.

The problem can be remedied as follows.
Write $d^4xd^2\theta =d^2z\cdot d^2\bar z d^2\theta$.
Then the factor  $d^2\bar z d^2\theta$ is in fact naturally defined
on any Kahler manifold (as should be clearer, along with some
subsequent claims,  in \S3), so the question
is to generalize $d^2z$.

The factor $d^2z$, or better $m \,d^2z$, including the factor of $m$ from
\immmo, can be interpreted as a holomorphic two-form on ${\bf R}^4$.
The generalization is immediate: on any Kahler manifold, pick a holomorphic
two-form $\omega$ and replace $m \,d^2z$ by $\omega$.
The generalization of \immmo\ to a general Kahler manifold is thus
$$\Delta L = -\int \omega\wedge d^2\bar z
 d^2\theta \Tr\Phi^2+{\rm h.c.} \eqn\bimmmo$$
We will verify in \S3 that this is of the form $\sum_a\alpha_aI(\Sigma_a)
+\{Q_1,\dots\}$.

\REF\donan{S. Donaldson, ``Irrationality And The $h$-Cobordism Conjecture,''
Jour. Diff. Geom. {\bf 26}  (1987) 141.}
At this stage, however, an important restriction arises.  A non-zero
holomorphic two-form $\omega$ exists not on arbitrary Kahler manifolds,
but only on those on which $H^{2,0}(M)\not= 0$.  So we are limited
to Kahler manifolds obeying that condition.  This is in fact
natural.  One of Donaldson's earliest results [\donan] was a sort of anomaly:
the topological correlation functions of the twisted $N=2$ theory
are in fact topological invariants only under a certain topological
condition (the self-dual part of $H^2(M)$ should have dimension bigger
than one); specialized to Kahler manifolds, the condition is precisely
$H^{2,0}(M)\not= 0$.  So the restriction required by our methods is in fact
the ``right'' one.\foot{There are two caveats here: (1)
There is one other exceptional case of a Kahler manifold, namely
${\bf CP}^2$, for which the Donaldson invariants are well-defined;
that is, the correlation functions are independent of the metric.
The construction of this paper does not apply and I have no reason to
expect the correlation functions to be given by similar formulas.
(2) Even when $H^{2,0}(M)=0$
and the correlation functions of the twisted theory are not topological
invariants, they still have very nice properties (which have been much
exploited mathematically) and there may be a useful way to study them
using physical techniques.}

\section{Some Simple Computations}
\def\bz{{\bf Z}}
\def\bbz{{\bf Z}_2{}'}
Now let us carry out some actual computations of the Donaldson
invariants for Kahler manifolds.
Since there is a mass gap, we can proceed much as before; but
there are a few subtleties.\foot{Apart from issues I discuss below,
one must consider the violation of
Lorentz invariance of the twisted theory
by the $N=1$ invariant bare mass term.  Lorentz invariance was
used in analyzing the two-forms to arrive at \cvovvo.
We will show in \S3 how to derive at \cvovvo, by a slightly lengthier
route, using properties that are preserved by the bare mass.}

For simplicity
we will specialize to gauge group $G=SU(2)$.  So the global symmetry group
is ${\bf Z}_4\times {\bf Z}_2{}'$, spontaneously broken to
$\bz_2\times \bbz$; as a result there are two vacuum states,
which we will call $|+\rangle$ and $|-\rangle$.
The sum over vacuum states in formulas such as
\cvovvo\ therefore involves two terms.
The constants $\eta$ and $\langle\co\rangle$ that appear in that
formula are odd under the broken symmetry, so $\eta_+=-\eta_-$
and $\langle\co\rangle_+=-\langle\co\rangle_-$.    We will
normalize the operators $\co$ and $I(\Sigma)$ so that
$\eta_\pm = \pm 1 $ and $\langle\co\rangle_\pm = \pm 2$.  By comparing
to special
cases of Donaldson invariants that have been computed mathematically,
one can verify that these normalizations agree with the standard
topological normalizations.

A subtlety arises in using the broken symmetry
to relate the contributions of the $|+\rangle$ and $|-\rangle$
vacua in formulas
such as \cvovvo.  The global symmetry group $\bz_4\times \bbz$
of the theory is a subgroup of the classical symmetry group consisting
of symmetries that are not explicitly broken by gauge instantons.
However, gravitational instantons -- in this case the effects
of working on a four-manifold
$M$ -- might induce additional symmetry breaking.  We should
remember that the $\bz_4$ symmetry is generated by
$$\alpha:~
\lambda\to i\lambda,\,\,\bar\lambda\to -i\bar \lambda.\eqn\hoffo$$
  In an instanton
field of instanton number
$k$, the net number of $\lambda$ minus $\bar\lambda$
zero modes  is, from the index theorem,
$$\Delta = 4k -3(1-h^{1,0}+h^{2,0}), \eqn\igaro$$
with $h^{p,q}$ the dimension of $H^{p,q}(M)$.
This can also be written in a way that makes sense for any
four-manifold (but might not be integral in general):
$$\Delta = 4k -{3\over 4}\left(\chi+\sigma\right)={1\over 2}\dim {\cal M}
, \eqn\figaro$$
with $\chi$ and $\sigma$ the Euler characteristic and the signature of $M$,
and ${\cal M}$ the moduli space of instantons.

Under the symmetry \hoffo, the integration measure for the fermion
zero modes transforms as $i^{-\Delta}$.  Therefore, the underlying
broken symmetry does not, as one might expect, cause the constants $C_+$ and
$C_-$ in formulas such as \cvovvo\ to be equal.  Rather, the relation is
$$C_-=i^{\Delta} C_+. \eqn\jupiter$$
Notice that the $k$-dependence drops out of this relation, because
the spontaneously broken symmetry has no anomaly under Yang-Mills instantons.
Of course, similar formulas can be worked out for groups
other than $SU(2)$ -- the difference being that there are not two
but $h$ vacua, each with its own constant $C$.

Using \figaro, we can also write
$$C_-=\exp\left(-{3i\pi\over 8}(\chi+\sigma)\right)C_+, \eqn\humble$$
showing that the relation between $C_-$ and $C_+$ is compatible with
the fact that each can be written as in \unicon.

\subsection{Hyper-Kahler Manifolds}

To work out actual formulas,
we consider first the special case of hyper-Kahler manifolds.  This
is the case in which the holomorphic two-form $\omega$, which is
responsible for the mass gap of the theory, has no zeroes, so the mass
gap can be applied most simply.

Moreover, for a hyper-Kahler manifold, the holonomy is not
$SU(2)_L\times U(1)_R$ (as for general Kahler manifolds)
but simply $SU(2)_L$.  This means that the twisting is trivial,
and the physical model coincides with the topological model.

Actually, there are only two cases of compact, non-singular
hyper-Kahler manifolds of dimension four: a four-torus ${\bf T}^4$,
and a $K3$ surface.
For ${\bf T}^4$, $h^{1,0}=2$, $h^{2,0}=1$, and $\Delta = 0$ modulo four,
so $C_-=C_+$ according
to \jupiter.  In fact, for the four-torus $\chi=\sigma=0$, and therefore
$C_+=C_-=1$, according to \unicon.
Hence the sum in \cvovvo\ takes the following form in this case:
$$\eqalign{
\left\langle \exp\left(\sum_a\alpha_a
I(\Sigma_a)+\lambda\co\right) \right\rangle
 = &\exp\left({1\over 2}\sum_{a,b}\alpha_a\alpha_b\#(\Sigma_a\cap \Sigma_b)
+2\lambda\right)\cr &+
\exp\left(-{1\over 2}\sum_{a,b}\alpha_a\alpha_b\#(\Sigma_a\cap \Sigma_b)
-2\lambda\right).\cr} \eqn\cooo$$
Unfortunately, mathematical computations of Donaldson invariants of
the four-torus have not been performed, so there is nothing to compare
\cooo\ to.  Incidentally, it is straightforward to extend \cooo\ to
include the additional observables associated with the odd dimensional
cohomology of the torus.

It remains to consider the case of $K3$.  For $K3$, $h^{1,0}=0$,
$h^{2,0}=1$, so from \jupiter, $C_-=-C_+$.  Therefore, \cooo\ is replaced
by
$$\eqalign{
\left\langle \exp\left(\sum_a\alpha_a
I(\Sigma_a)+\lambda\co\right) \right\rangle
 =&C
\left(\exp\left({1\over 2}\sum_{a,b}\alpha_a\alpha_b\#(\Sigma_a\cap \Sigma_b)
+2\lambda\right)\right.\cr &-\left.
\exp\left(-{1\over 2}\sum_{a,b}\alpha_a\alpha_b\#(\Sigma_a\cap \Sigma_b)
-2\lambda\right) \right),\cr}\eqn\booo$$
with an unknown constant $C$.  According to computations of [\kthree],
this formula is in fact correct with $C=1/4$.

Chiral symmetry breaking is the reason that
the right hand sides of both \cooo\ and \booo\ are given by not a single
exponential but
a sum of two exponentials, with a relative plus sign in one case
and minus sign in the other case.  These signs play the following role.
For the four-torus, for dimensional grounds explained in \ippso,
a correlation function
$$\left\langle \co(x_1)\dots \co(x_r)I(\Sigma_1)\dots I(\Sigma_s)
\right\rangle\eqn\moon$$
vanishes unless $4r+2s$ is divisible by 8.  For the $K3$ surface,
the requirement is that $4r+2s$ should be congruent to 4 modulo 8.
A single exponential of the form in \cooo\ or \booo\ would not obey
these requirements, but the given sums of two exponential with
plus or minus signs do obey them.  Thus (given the
mass gap of the $N=1$ theory and the other constraints) chiral symmetry
breaking is essential in making sense out of Donaldson theory.  This
fact (of which we will see subtler versions later, involving the
canonical divisor) is a sort of global generalization of old insights
[\russians] concerning supersymmetric Yang-Mills theory on ${\bf R}^4$.

\section{General Kahler Manifolds}

Now we consider the case of a general Kahler manifold $M$, with
$H^{2,0}\not= 0$.  The important
new phenomena spring from the fact that (if $M$ is not hyper-Kahler)
the canonical class of $M$ is non-trivial.  This
implies that the mass term
$$\Delta L=-\int_M\omega\wedge d^2\bar z d^2\theta\Tr \Phi^2-{\rm h.c.}
\eqn\harf$$
is not everywhere non-zero.  The $\Phi$ mass vanishes where $\omega$
does; and $\omega$ vanishes on a divisor $C$ representing the canonical
class of $M$.  For simplicity we will assume -- at least initially --
that $C$ is the union of smooth, disjoint Riemann surfaces
$C_y$ along which $\omega$ has simple zeroes.

The local model that we reduce to by scaling up the metric of $M$
is now the following.  Each $C_y$ becomes a flat complex dimension
one submanifold of a flat ${\bf R}^4\cong {\bf C}^2$.
One might think of the $C_y$ as world-sheets of cosmic strings.
The strings in question are similar to ``global strings'' (their
existence hinges on the vanishing of a gauge-invariant field $\omega$)
rather than ``local strings'' (in the core of which there is an enhanced
unbroken gauge symmetry).

The most important phenomenon associated with this particular
type of cosmic string is that it can capture fermion zero modes.
In fact, pick complex coordinates $z_1,z_2$ on ${\bf R}^4$,
and suppose that $\omega= z_1 dz_1\wedge dz_2$, with a simple
zero on the Riemann surface $C$ defined by $z_1=0$.
Any example (with non-singular $C_y$) reduces locally to this example
when the metric is scaled up.

In this example, the mass term of the
$\Phi$ multiplet
is proportional to $z_1$, so it changes in phase by $2\pi$ in making
a small circuit around the ``core of the string'' at $z_1=0$.  This
is the standard situation in which fermion zero modes
are trapped near the core of a string.  In fact, the Dirac
equation for the motion of the $\psi$ field (the fermionic part
of the $\Phi$ multiplet) in the $z_1$ plane
has a normalizable zero mode, as we will discuss in more detail in \S3.
When one allows
for the $z_2$ dependence of the $\psi$ wave-function, this ``zero
mode'' becomes a two-dimensional quantum field, propagating on the
string world-sheet, that is, the
$z_2$ plane.  We will call this two dimensional quantum field $\psi'$.
$\psi'$ is a spin one-half fermion in the adjoint representation
of $G$ (since those are the quantum numbers of $\psi$); it has
definite chirality, which we will conventionally call left-handed,
since the zero mode in the $z_1$ plane has definite chirality.

In explaining the origin of the effective two-dimensional field
$\psi'$, the gauge couplings were not essential; the crucial zero
mode develops because of the topology of the ``Higgs field'' $\omega$.
When we consider the gauge fields, a new consideration arises:
the coupling of gauge fields to two-dimensional chiral fermions $\psi'$
is anomalous.
Since the underlying four-dimensional theory is anomaly-free, there
must be additional fields trapped along the string that cancel the
anomaly.  To understand the details, one would need to know something
about how the strongly coupled $N=1$ supersymmetric Yang-Mills theory
behaves near the core of the string.  This appears out of reach at present.
But somehow, extra modes must be trapped along the string in such
a way that the effective theory along the string is a gauge-invariant,
anomaly-free theory that we will call the cosmic string theory.
It can be seen that the two-dimensional
cosmic string theory has $(0,2)$ supersymmetry, but we will not exploit that.

\section{Symmetries And Vacuum Structure Of The Cosmic String Theory}

Since we do not even know the fields and Lagrangian of the cosmic string
theory, how can we proceed?  We will make what reasonable conjectures
we can based on the symmetries.

As we recall, the bulk theory had a $\bz_{2h}\times \bbz$ global
symmetry, which we believe is spontaneously broken in bulk to
$\bz_2\times \bbz$.  Along the string, the question of symmetry
breaking must be re-examined; further symmetry breaking might
occur in the core of the string.

In bulk, $\bbz$ couples only to particles of positive bare mass,
and it was implausible that it would be spontaneously broken.
Along the string, however, $\bbz$ couples to (and in fact only to)
the massless chiral
fermion $\psi'$, so in the cosmic string theory $\bbz$ is in fact a chiral
symmetry.  It is very plausible that such a chiral symmetry would
be spontaneously broken.

On the other hand,
a diagonal subgroup $\bz_2{}''\subset\bz_2\times \bbz$ is generated
by the operator $(-1)^F$ that counts fermions modulo two; this is equivalent
to a $2\pi$ rotation of the cosmic string, and cannot be spontaneously
broken if rotation invariance is unbroken.

\def\bp{+}
\def\bm{-}
We will make for
the cosmic string theory
the same sort of assumption that we made in bulk.  We will assume
that the cosmic string theory has a mass gap and that the only vacuum
degeneracy comes from symmetry breaking.  (Thus, in particular, the
vacuum degeneracy can be measured by local operators.)
For the pattern of symmetry breaking, I will assume that
$\bz_2\times \bbz$
is spontaneously broken to $\bz_2{}''$ -- the largest unbroken symmetry
that permits fermion masses.  Thus -- regardless of the gauge group --
I postulate that the cosmic string theory has two vacuum states
(for a given choice of the vacuum in the bulk four-dimensional theory).

Hitherto we have considered the symmetries of the cosmic string theory
which are (i) not explicitly broken by Yang-Mills instantons; (ii)
not spontaneously broken in bulk.
As in the bulk theory, for instance the discussion of \jupiter,
we must in addition worry about whether there is further explicit
breaking of symmetries by gravitational instantons.  What this means
concretely is as follows.  We consider the actually relevant
case in which the string
world sheet is not the $z_2$ plane but a compact Riemann surface $C_y$;
we need to know whether the path integral measure along $C_y$ for the
effective two-dimensional theory is even or odd under the broken
$\bbz$ symmetry.

As in the bulk discussion,
the question comes down to counting modulo two the $\psi'$ zero modes.
If the number of zero modes is even,  the measure is even under $\bbz$;
if the number of zero modes is odd, the measure is odd.  The number
of zero modes modulo two is a topological invariant, independent of the
gauge fields (and all fields other than $\psi'$), so the problem
can be treated as a problem of free fermions along $C_y$.

Let $d$ be the dimension of the gauge group.
The $\psi'$ field has $d$ components; if it is treated as  a free field,
the number
of zero modes is $d$ times what it would be for a single chiral fermion.
If $d$ is even, there is no further issue; the number of zero modes,
and the path integral measure, are even.  More subtle is the case of an
odd dimensional group, such as $SU(2)$.  The result then depends on the
spin structure. The number of zero modes is even or odd for chiral
fermions coupled to a so-called even or odd spin structure.
Let $\epsilon_y$
be $0$ or $1$ depending on whether the spin structure is even or odd.
Then the number of zero modes along $C_y$ is $d\epsilon_y$, modulo two,
so the path integral measure transforms under the broken $\bbz$ symmetry
by a factor of
$$t_y =(-1)^{d\epsilon_y}.           \eqn\burry$$

The spin structure
that arises here cannot be chosen arbitrarily but must be deduced from
the underlying four-dimensional theory.
This will be done in \S3.3 with the following result.
If $C_y\subset M$ is a smooth Riemann surface defined by a first order zero
of a holomorphic two-form $\omega$, then the inverse of
the normal bundle $N_y$ to
$C_y$ in $M$
can be interpreted as a spin bundle of $C_y$.  This (or actually
its complex conjugate) is the spin bundle
of which $\psi'$ is a section; its evenness or oddness determines
(if $d$ is odd) whether $\epsilon_y$ is 0 or 1 and so
whether the measure is even or odd under the broken
symmetry.

\subsection{Computations}

Now we come to the computational stage.
Imitating our considerations in bulk, the first step is
to analyze the partition function of the cosmic string theory,
with a given choice of the bulk vacuum state $|+\rangle$ or $|-\rangle$
 and a given
refinement to include the vacuum structure along the string.
We will label the four vacuum states in the vicinity of a particular
string as $|++\rangle, \,\,|+-\rangle,\,\,|-+\rangle,\,\,$
and $|--\rangle$, with the first sign indicating the ambient behavior
in bulk and the second sign indicating the behavior near the string.

As in the derivation of \uddu, the assumption of a mass gap implies
that  the partition function of the cosmic string theory on a Riemann surface
$C$ must be of the form
$$\langle 1\rangle_C=\exp\left(-\int_Cd^2x \sqrt g W(x)\right) \eqn\huffor$$
with $W(x)$ a local operator constructed from the Riemannian geometry.
The only topological invariant of a Riemann surface that can be constructed
in this way (indeed, its only topological invariant at all) is the Euler
characteristic $\chi(C)$.  It would appear, therefore,
that the partition function of the four-dimensional theory, allowing
for contributions from all of the $C_y$, would have a factor
$$\prod_y\exp\left(-w\chi(C_y)\right)=\exp\left(-w\sum_y\chi(C_y)\right),
 \eqn\hrr$$
with some universal coefficient $w$.  However, using standard facts
in complex geometry, $\sum_y\chi(C_y)$ is a linear combination of
$\sigma(M)$ and $\chi(M)$, so this factor can be absorbed in adjusting
the constants $a$ and $b $ in \uddu.

We also have to consider the local operators that arise in the cosmic
string theory.  The local operator $\co(x)$ of the original four-dimensional
theory can be inserted at points disjoint from the $C_y$,
and so does not contribute to the cosmic string theory.  However,
the other operator $I(\Sigma)$ is constructed as an integral over
a two-manifold $\Sigma$:
$$I(\Sigma)=\int_\Sigma Z_{mn}d\sigma^{mn}.    \eqn\uvv$$
Topologically, intersections of $\Sigma$ with the $C_y$ may be
unavoidable; they can, however, be taken to be transverse.

We recall that when the metric of $M$ is scaled up, the bulk contribution
to the
integral in \uvv\ can be discarded.  In our earlier analysis, we obtained
local contributions at points of intersections of $\Sigma$'s.  Similarly,
local contributions will arise at intersections of the $\Sigma_a$ with
the $C_y$.  At every intersection point $P $ of $\Sigma_a$
and $C_y$, the integral \uvv\ will contribute a local operator $V(P)$
in the cosmic string theory -- times a sign depending on the relative
orientation of $\Sigma_a$ and $C_y$.
(As in the discussion of intersections
of $\Sigma$'s, such a sign must appear, since $I(\Sigma)$ is odd under
reversal of orientation of $\Sigma$.)  So, if $\#(\Sigma\cap C_y)$
is the algebraic intersection number of $\Sigma$ and $C_y$, then
one can make the replacement
$$I(\Sigma)\to \sum_y\#(\Sigma\cap C_y) V_y +{\rm terms~involving
{}~intersections~of~}\Sigma'{\rm s}.  \eqn\jury$$
Here $V_y$ denotes the operator $V$ inserted on $C_y$.

Inside correlation functions, the scalar operator $V$ of the cosmic
string theory can be replaced by its vacuum expectation value
-- like any local operator in a theory with a mass gap.
We must work out the expectation value of $V$ in the various vacuum
states, which by our hypotheses are related by spontaneously broken
symmetries.  Let us therefore describe the symmetry and vacuum structure
precisely.  Let $\alpha$ and $\beta$ be the generators $\alpha:
\lambda\to i\lambda,\Phi\to\Phi$ and $\beta:\lambda\to\lambda,\Phi\to
-\Phi$ of the $\bz_4$ and $\bbz$ factors, respectively, of the global
symmetry group
$\bz_4\times \bbz$.  Then, as we will see in \S3.3,
$$\alpha V = i V,  ~~~~ \beta V=-V. \eqn\nurdox$$
In bulk there are two vacua, $|+\rangle$ and $|-\rangle$, with
$$\alpha |+\rangle =|-\rangle,~~~\alpha|-\rangle = |+\rangle.\eqn\ufff$$
Allowing for the behavior near the string, these two states bifurcate
into four, as described earlier; we can choose the labeling of the vacua
so that
$$\eqalign{\alpha|++\rangle&=|-+\rangle \cr
           \alpha|-+\rangle & =|+-\rangle \cr
           \alpha|+-\rangle & = |--\rangle \cr
           \alpha|--\rangle & = |++\rangle .\cr }\eqn\jocco$$
The action of $\beta$ need not be separately given as $\alpha^2\beta$
is unbroken and hence $\beta$ acts on the vacua as $\alpha^2$.
If therefore the expectation value of $V$ in the state $|+\bp\rangle$
is $v$, then altogether
$$\eqalign{\langle V\rangle_{+\bp} & = v \cr
           \langle V\rangle_{-\bp} & = -iv \cr
           \langle V\rangle_{+\bm} & = -v \cr
           \langle V\rangle_{-\bm} & = iv.\cr}\eqn\guffrey$$

\subsection{Final Evaluation}

Finally, then, to obtain the topological correlation functions
for Kahler manifolds obeying our assumptions, one must sum over
vacuum states in bulk and along the $C_y$.  Relative phases
in the contributions of different vacua come from \jupiter\ and \burry.
The parameters $\Delta$ and $t_y$ appearing in those
equations are not independent, but are related by
$$\Delta +\sum_y\epsilon_y\cong 0~{\rm modulo}~2, \eqn\rippo$$
(with $t_y=(-1)^{\epsilon_y}$), as we will explain in \S3.4.

To compute correlation functions of $I(\Sigma)$'s,
one determines the cosmic string contributions
as in \jury, and replaces $V$ by
its expectation values just given.

In writing down the final result, I will adjust the constants $a$ and
$b$ in \uddu\ and the constant
$v$ in \guffrey\ to
values chosen to agree with special cases of mathematical
computations.  The requisite values turn out to be $a=(7/4)\ln 2$, $b=
(11/4)\ln 2$,
and $v=1$.
The constants $\eta$ and $\langle\co\rangle$ were
similarly adjusted earlier, so we have fixed five universal constants
by comparing to particular calculations of Donaldson invariants.

We want the general formula for
$$\left\langle\exp\left(\sum_a\alpha_aI(\Sigma_a)+\lambda \co\right)
\right\rangle.         \eqn\murkky$$
It is convenient to set
$$\phi_y=\sum_a\alpha_a\#(\Sigma_a\cap C_y).\eqn\convset$$
According to \jury, the cosmic string contribution
to the exponent in \murkky\ is $\sum_y\phi_yV_y$.  Replacing
this with its expectation values in the various states, we see that
the $|+\rangle$ vacuum, with its various bifurcations along the $C_y$,
contributes
$$2^{{1\over 4}\left(7\chi + 11\sigma\right)}
\exp\left(\sum_{a,b}{\alpha_a\alpha_b\over 2}\#(\Sigma_a\cap \Sigma_b)
+2\lambda \right)\prod_y\left(e^{\phi_y}+t_ye^{-\phi_y}\right).
       \eqn\onvset$$
The $|-\rangle$ vacuum,
with its various bifurcations, contributes the chiral
transform of this, or
$$i^{\Delta}
2^{{1\over 4}\left(7\chi + 11\sigma\right)}
\exp\left(-\sum_{a,b}{\alpha_a\alpha_b\over 2}\#(\Sigma_a\cap \Sigma_b)
-2\lambda \right)\prod_{y}\left(e^{-i\phi_y}+t_ye^{i\phi_y}\right).
       \eqn\onovset$$
As a check, note that this is real by virtue of \rippo.

The final formula for the $SU(2)$ Donaldson invariants of Kahler manifolds
(with $H^{2,0}(M)\not= 0$ and possessing a holomorphic two-form
whose zero set is a union of smooth disjoint complex curves of multiplicity
one) is obtained by summing these contributions and is
$$\eqalign{
&\left\langle\exp \left(\sum_a\alpha_aI(\Sigma_a)+\lambda \co\right)
\right\rangle_T \cr
=&~~~ 2^{1+{1\over 4}\left(7\chi + 11\sigma\right)}
\exp\left(\sum_{a,b}{\alpha_a\alpha_b\over 2}\#(\Sigma_a\cap \Sigma_b)
+2\lambda \right)\prod_{y}\left(e^{\phi_y}+t_ye^{-\phi_y}\right)\cr
&~~~+i^{\Delta} 2^{1+{1\over 4}\left(7\chi + 11\sigma\right)}
\exp\left(-\sum_{a,b}{\alpha_a\alpha_b\over 2}\#(\Sigma_a\cap \Sigma_b)
-2\lambda \right)\prod_{y}\left(e^{-i\phi_y}+t_ye^{i\phi_y}\right).
  \cr}         \eqn\nurkky$$
Here (as implied by the symbol $\langle~~~\rangle_T$)
we use the normalization of topologists, and so
include a factor of 2 -- the order of the center of
$SU(2)$ -- from \smallcorrection.

Here is a check on \nurkky.  Using \figaro\ and \ippso,
the dimension of the moduli space of
Yang-Mills instantons is $2\Delta$, and the correlation function
$$\left\langle
\co(x_1)\dots \co(x_r) I(\Sigma_1)\dots I(\Sigma_s)\right\rangle
\eqn\bufus$$
is zero unless
$$ 4r+2s= 2\Delta. \eqn\knon$$
Therefore, if we consider the $\alpha_a$ to be of
degree 2 and $\lambda$ to be of degree 4, then the power series
expansion of \nurkky\ should have only terms of degree congruent to
$2\Delta$ modulo 8.  It is easy to verify this.

Unlike the naive formula \cvovvo\ that
would arise if the $N=2$ theory had a mass gap, the formula
is not manifestly a topological (or differentiable) invariant.
Indeed, the formula appears to depend on the complex structure of $M$
(which determines the canonical divisor and thus the ultimately
the $\phi_y$).  The metric independence of the underlying twisted
$N=2$ theory ensures, however, that the formula really only depends
on the structure of $M$ as a differentiable manifold.  The fact
that the formula is a differentiable invariant but not obviously
so is responsible for its mathematical interest.

Thus, a fairly typical mathematical application of
Donaldson theory is as follows.  In many instances one can construct
pairs of Kahler four-manifolds $M_1$ and $M_2$ which have the
same values of the obvious topological invariants but for which the
formulas \nurkky\ do not coincide.  Then one can infer that $M_1$
and $M_2$ are in fact not isomorphic as smooth manifolds.  In
practice, the data present in \nurkky\ beyond the classical
invariants are the cohomology classes of the connected components $C_y$
of the canonical divisor and the types (even or odd) of the normal
bundles to the $C_y$; $M_1$ and $M_2$ must have the same values of
this data if they are to coincide as smooth four-manifolds.

\subsection{Simple Type Condition And Comparison To Known Formulas}

A four-manifold is said to be of simple type if
$f=\langle\exp(\sum\alpha_aI(\Sigma_a)+\lambda \co)\rangle$
obeys the equation $\partial^2f/\partial\lambda^2=4f$.
\nurkky\  evidently implies that Kahler manifolds obeying our
assumptions are of simple type.
The derivation of \cvor\ shows that the simple type condition
is really a corollary of having a mass gap.  Thus, the assumption
that $H^{2,0}(M)\not= 0$, which permitted us to reduce to a theory
that has a mass gap, was essential; we would not necessarily expect
${\bf CP}^2$, for instance, to obey a condition similar to the
simple type condition.

In deriving \nurkky\ we assumed also that there exists
a holomorphic two-form $\omega$ that vanishes precisely on a union
of disjoint smooth curves.
This is not relevant to the simple type condition, since regardless
of the structure of the canonical divisor, one can assume that
$\co$ is inserted at points disjoint from it.  Therefore, only
the bulk vacuum structure is relevant in determining the dependence
of $f$ on $\lambda$, so (at a physical level of rigor) all Kahler
manifolds with $H^{2,0}\not= 0$ are of simple type.

According to Kronheimer and Mrowka [\kronheimer], on a four-manifold
of simple type, $f$,
as a function of the $\alpha_a$, is a Gaussian times a sum
of exponentials.  (In fact,  they make more precise claims.)
This was a clue suggesting the role of a mass gap
and $N=1$ supersymmetry, and \nurkky\ has the expected properties.
We will say more about this presently.

\section{More General Kahler Manifolds}

In general, we should consider the case of a Kahler manifold
$M$ with $H^{2,0}\not= 0$, but with a canonical divisor that
does not have the simplifying properties that we have so far assumed.
The zeroes of a holomorphic two-form $\omega$ will be a union
of connected curves $C_y$; but -- even if $\omega$ is chosen
generically -- the $C_y$ may not be smooth, and $\omega$ may vanish
along components of $C_y$ with multiplicity greater than one.

No matter how badly behaved the canonical divisor may be,
the mass gap ensures that the contribution of a given $C_y$
depends only on the local structure of $C_y$ and its intersections
with the $\Sigma_a$.  We recall that the $C_y$ were defined as the
connected components of the zero locus of $\omega$.  Each $C_y$
may in turn be a union of irreducible (but perhaps singular)
Riemann surfaces $C_{y,\beta}$ (which intersect each other, perhaps
at their singularities).  The only homology invariants in the intersections
of the $\Sigma_a$ and the $C_y$  are the algebraic intersection
numbers $\#(\Sigma_a\cap C_{y,\beta})$, and the mass gap
permits the correlation functions to depend on the natural
generalization of \convset:
$$\phi_{y,\beta}=\sum_a\alpha_a~\#(\Sigma_a\cap C_{y,\beta}).\eqn\onvset$$

The possible generalization of \nurkky\ allowed by the mass gap is
$$\eqalign{
&\left\langle\exp \left(\sum_a\alpha_aI(\Sigma_a)+\lambda \co\right)
\right\rangle_T \cr
=&~~~ 2^{1+{1\over 4}\left(7\chi + 11\sigma\right)}
\exp\left(\sum_{a,b}{\alpha_a\alpha_b\over 2}\#(\Sigma_a\cap \Sigma_b)
+2\lambda \right)\prod_{y}F_y(\phi_{y,\beta})\cr
&~~~+i^{\Delta} 2^{1+{1\over 4}\left(7\chi + 11\sigma\right)}
\exp\left(-\sum_{a,b}{\alpha_a\alpha_b\over 2}\#(\Sigma_a\cap \Sigma_b)
-2\lambda \right)\prod_yF_y(-i\phi_{y,\beta}),  \cr}         \eqn\urkky$$
where $F_y(\phi_{y,\beta})$ is a universal function, depending only on
the local structure of $\omega$ near $C_y$.
$F_y$ is a function of all the $\phi_{y,\beta}$ for $C_{y,\beta}$
a component of $C_y$.

We can be somewhat more precise about the nature of the function
$F_y$.  Let us focus on a particular component $C_y$ of the canonical
divisor; we will simply call it $C$, suppressing the $y$ index.
Let $C_\beta$, $\beta=1\dots w$
 be the irreducible components of $C$, and deleting
the $y$ index let $\phi_\beta=\sum_a\alpha_a\#(\Sigma_a\cap C_\beta)$.
Each $C_\beta$ has a multiplicity $n_\beta$ (the order of vanishing
of the two-form $\omega$ in approaching a generic point of $C_\beta$).
The vacuum state $|+\rangle$ of the bulk theory bifurcates along $C_\beta$
to a number of vacuum states
$|\rho_{i_\beta}\rangle$, where $i_\beta$
runs over a finite set $S_\beta$.
According to our assumptions above, if $n_\beta
=1$, then  $S_\beta$ is a set with two elements.
In what follows, I will make no assumption about the number of
vacua for $n_\beta>1$, but I will assume that the cosmic string
theory has a mass gap and a
finite vacuum degeneracy (which can be detected by local operators)
for any value of $n_\beta$.
\foot{In fact, a reasonable
guess is that in general the cosmic string theory has
$2^{n_\beta}$ vacua.  The logic behind this is that
a component $C_0$ with a given multiplicity $n_0$ arises
from a two-form with an $n_0{}^{th}$ order zero along a complex
curve, say $\omega=z_1{}^{n_0}dz_1\wedge dz_2$.  Such a two-form
can be deformed to  a two-form with simple zeroes on $n_0$ distinct
curves, say $\omega'=\prod_{i=1}^{n_0}(z_1-a_i)dz_1\wedge dz_2$.
After the perturbation, there are $2^{n_0}$ vacua, coming from a
two-fold degeneracy on each of the $n_0$ components.  If the mass gap
holds uniformly during the process of deformation from $\omega$ to
$\omega'$, then $2^{n_0}$ is also the number of vacua prior to the
deformation.}
A vacuum state of the cosmic string theory can be labeled by
a $w$-plet $\vec \rho=(\rho_{i_1},\dots,\rho_{i_w})$.  Indeed, once
the generic behavior along the $C_\beta$ is given, all that remain
are finitely many isolated
singular points; no further issues of infinite volume
and vacuum degeneracy arise in discussing
the behavior near such isolated points.

The function $F$ is simply the expectation in the cosmic string theory
associated with $C$ of the operator $\exp(\sum_a\alpha_aI(\Sigma_a))$.
As in \jury, the exponent can be replaced by $\sum_\beta\phi_\beta V_\beta$
where $V_\beta$ is the operator $V$ inserted at a generic point of $C_\beta$
and $\phi_\beta=\sum_a\alpha_a\#(\Sigma_a\cap C_\beta)$.  In sum,
we must calculate the expectation value of the operator
$\exp(\sum_\beta\phi_\beta V_\beta)$
in the cosmic string theory.  Scaling up the
metric, this reduces in the usual way to a sum over the contributions
of vacuum states,
$$F=
\sum_{\vec\rho}a_{\vec\rho}\left\langle\exp\left(\sum_\beta\phi_\beta
V_\beta\right)\right\rangle_{\vec\rho}, \eqn\undero$$
with $a_{\vec\rho}$ being a constant measuring the amplitude with which
the vacuum $\vec\rho$ appears.  (Apart from the sort of
effects considered in deriving
\burry\ and \huffor, the $a_{\vec \rho}$ might receive
contributions involving the structure of the singularities of $C$. The
singularities, being isolated, do not affect the bulk or vacuum structure
along $C$, but might contribute volume-independent multiplicative factors.)
This sum can be factored as a product of contributions from the
individual components $C_\beta$:
$$F=\sum_{\vec\rho=(\rho_{i_1},\dots,\rho_{i_w})}
a_{\vec\rho}\prod_{\beta=1}^w\exp\left(\phi_\beta\langle V_\beta
\rangle_{i_\beta}\right).\eqn\piccolo$$
Here $\langle V_\beta\rangle_{i_\beta}$
is the expectation value of $V_\beta$
in the vacuum $\rho_{i_\beta}$.  This is our final result.

So in particular, $F$ is a sum of exponential
functions of the $\phi_\beta$.  In view of how $F$ enters in \urkky,
this is in fact the general structure proved by Kronheimer and Mrowka
[\kronheimer] for four-manifolds of simple type.  We actually
have somewhat more precise information about the exponents that
arise in $F$.   (These exponents
are the ``simple classes'' in the language
of Kronheimer and Mrowka.)  For $n_\beta=1$, the expectation
value $\langle V_\beta
\rangle_{i_\beta}$ equals $1$ or $-1$, with our
normalization above.  For general $n_\beta$, the possible
values of $\langle V_\beta\rangle_{{i_\beta}}$
depend only on $n_\beta$.  The simple classes
are of the form
$$\sum_{y,\beta} \langle V_\beta\rangle_{{i_\beta}}
C_{y,\beta},\eqn\gufno$$
with $i_\beta\in S_\beta$ and so are determined by the $n_\beta$.

If one is willing to make the assumption of the last footnote,
one can be more precise: the possible values of $\langle V\rangle$
are of the
form $\pm 1\pm 1\dots \pm 1$ with a sum of $n_\beta $ terms, which
are the contributions of the $n_\beta$ ``ordinary'' cosmic
strings into which the multiplicity $n_\beta$ string can be deformed.
The expectation values are thus $n_{\beta},n_{\beta}-2,n_\beta-4,\dots,
-n_\beta$, and the simple classes are thus
$$\sum_{y,\beta}a_{y,\beta} C_{y,\beta}, \eqn\hdind$$
where the $a_{y,\beta}$ are integers no greater in absolute value than
the multiplicity $n_{y,\beta}$ and congruent to $n_{y,\beta}$ modulo two.
There is actually some additional evidence for this formula:
by considering the special case of an elliptic surface with
a multiple fiber, and comparing to a conjecture presented at the end
of [\kronheimer], one can see that if their conjecture is true,
the expectation values and simple classes are as just stated.
(The multiple fiber can have an arbitrary $n_\beta$, so a knowledge
of the possible values of $\langle V_\beta\rangle$ in this case
determine the general structure of the simple classes.)
It may be that the list of simple classes is shorter than just
claimed, as some of the $a_{\vec \rho}$ may vanish.

There remains the question of whether the constants $a_{\vec\rho}$
of equation \undero\ contain any deep information about four-manifolds.
I expect
the answer to be negative because the function $F$ is invariant
under a rather crude equivalence relation.  Because of the mass gap,
in studying the cosmic string theory near $C$,
one can restrict to a small open neighborhood of $U$; because of the
underlying topological invariance, one can then deform the complex
structure of $U$ and the two-form $\omega$.  I expect that using
this relation one can eliminate the singularities of $C$ and reduce
to the case that a connected component $C$
of the canonical divisor is a smooth curve
of multiplicity 1 or a
genus one curve of multiplicity $n$ with normal bundle of order $n$.
In the former case, from our above results,
$F(\phi) = 2\cosh \phi$
or $2\sinh\phi$ depending on the type of the normal bundle; in the
latter case, according to the conjecture of Kronheimer and Mrowka,
$F(\phi)=\sinh(n+1)\phi/\sinh\phi$.

\subsection{Kahler Manifolds With $\pi_1\not= 0$}

To complete the discussion of Kahler manifolds, it remains to discuss
the case of a Kahler manifold $M$ with $\pi_1(M)\not= 0$.

Actually, the fundamental group played no particular role in
our considerations (given that we assume that there are no unbroken
and unconfined gauge symmetries), and as long as one restricts to
the operators $\co(x)$ and $I(\Sigma)$, the correlation functions
are still given by the above formulas.  The only real
novelty is the following.
When the odd Betti numbers of $M$ are non-zero,  one can
consider additional operators $I(T)=\int_T\co^{(r)}$, with $T$ an
$r$-cycle in $M$ for $r=1$ or $r=3$.

Correlation functions of the new operators can be analyzed
with the same methods used for $I(\Sigma)$ and $\co$.
The mass gap means that the correlation functions
depend only on intersections of the various cycles with each other
and with the canonical divisor.  Just as in our discussion of two-cycles,
the intersections that do not involve the canonical divisor detect
only the classical intersection ring of $M$.  The interesting information
comes from intersections involving the canonical divisor $C$;
the only such intersection involving odd-dimensional cycles is
$T\cap T'\cap C$ with $T$ and $T'$ being three-cycles.
Since $\Sigma=T\cap T'$ is a two-dimensional class,
we are again studying intersections $\Sigma\cap C$. The only novelty
might be that the operator $V$  of the above discussion would
be replaced by another operator of the cosmic string theory, but I doubt
that anything essentially new can be detected in this way.

\section{Other Gauge Groups}

Before discussing other gauge groups in general, let us discuss
one particularly simple case: $G'=SO(3)$.  Because $G'$ is locally
isomorphic to $G=SU(2)$, all local considerations are unchanged.
The symmetries, the symmetry breaking pattern, the
vacuum structure, and the constants $a,b,\eta,\langle \co\rangle$,
and $\langle V\rangle$, are all determined by the structure
on ${\bf R}^4$ (perhaps in the presence of an infinite straight cosmic
string) and are unchanged in going from $SU(2)$ to $SO(3)$.
Only a few global factors need modification.

The only difference between the $SO(3)$ theory and the $SU(2)$
theory is that there are $SO(3)$ bundles that cannot be derived
from $SU(2)$ bundles.  Given an $SO(3)$ bundle $E$, the obstruction
to lifting its structure group to $SU(2)$ is the second Stiefel-Whitney
class $w_2(E)$.  Before proceeding, we must formulate precisely
what we mean by the $SO(3)$ theory.  It is natural to define
an $SO(3)$ theory by summing over all isomorphism classes of
$SO(3)$ bundles,
but to get the sharpest mathematical
formulas, I will instead fix the second
Stieffel-Whitney class and sum over all bundles with a given
value of $w_2(E)$.  Also, I will take the gauge group to be the
``same'' as the $SU(2)$ gauge group in the sense of only allowing
gauge transformations which, along
any path on which the bundle $E$ is trivialized,
lift to $SU(2)$ gauge transformations. (Otherwise we would have
to divide by an extra factor equal to the number of components
of the group of $SO(3)$ gauge transformations; this is
$\# H^1(M,{\bf Z}_2)$.  Also, as $SO(3)$ has no center, the factor
of 2 in \smallcorrection\ in comparing different normalizations
would not appear.)

The remaining global issues that entered the $SU(2)$ analysis
involved the action of the global symmetries on the space of vacua.
Since the $SO(3)$ theory coincides with the $SU(2)$ theory on
${\bf R}^4$, it has after perturbing by the bare mass the familiar
global symmetry group $\bz_4\times \bbz$.  However, on a four-manifold
$M$, the path integral measure may transform differently.
As we recall, the transformation of the path integral measure
depends on index theorems that determine the number of fermion
zero modes both in bulk and along the cosmic strings.

If $M$ is simply-connected,  $w_2(E)$
can be lifted to an integral cohomology class (which we will also call
$w_2(E)$) which is well-defined modulo two; hence $w_2(E)^2$ is well-defined
modulo four.  Even if $M$ is not simply-connected, one can still
make sense of $w_2(E)^2$ modulo four, as noted in [\kd, p. 41].
The expression $c_2(E)=\int_M\Tr F\wedge F/8\pi^2$
(with $\Tr$ an invariant quadratic form on the Lie algebra
normalized to agree with the trace in the two-dimensional representation
of $SU(2)$) is
not necessarily an integer if $E$ cannot be derived from an $SU(2)$
bundle.  The relation is rather
$$c_2(E)\cong -{w_2(E)^2\over 4}~{\rm modulo}~{\bf Z}.  \eqn\cnon$$
Consequently, the formula \igaro\ for the net number of $\lambda$
zero modes is modified to
$$\Delta_E= 4k-{w_2(E)^2}-3(1-h^{1,0}+h^{2,0}), \eqn\urphy$$
with $k\in {\bf Z}$.  The path integral measure therefore transforms
under the generator $\alpha$ of $\bz_4\times \bbz$ not as $i^\Delta$
but as $i^{\Delta_E}$, with
$$\Delta_E=\Delta-w_2(E)^2.        \eqn\furphy$$

Now let us consider what happens along a cosmic string component
$C_y$.  Let ${\cal S}$ be the relevant spin bundle along $C_y$.
The number of fermion zero modes modulo two is a topological
invariant of the bundle ${\cal S}\otimes E$.  If $w_2(E)$ is
zero when restricted to $C_y$, then we can trivialize $E$
along $C_y$; then ${\cal S}\otimes E\cong {\cal S}\oplus {\cal S}
\oplus {\cal S}$.  The number of zero modes of fermions coupled
to ${\cal S}\otimes E$
is hence congruent modulo two to the number $\epsilon_y$ of zero
modes of fermions coupled to ${\cal S}$.  Hence the transformation
law of the measure is $t_{y,E}=(-1)^{\epsilon_y}$ if $(w_2(E),C_y)=0$.
On the other hand, if $(w_2(E),C_y)\not= 0$, then we can take
$E$ restricted to $C_y$ to be ${\cal O}\oplus {\cal O}(n)\oplus
{\cal O}(-n)$, where ${\cal O}(n)$ is any line bundle of odd degree
$n$ and ${\cal O}(-n)$ is its inverse.  So ${\cal S}\otimes E
\cong {\cal S}\oplus {\cal S}(n)\oplus {\cal S}(-n)$; for large positive
$n$, the number of zero modes for fermions coupled to those three
line bundles are $\epsilon_y,n$, and 0.  Since $n$ is odd, this
is congruent to $\epsilon_y+1$ modulo two, so the measure
transforms as $t_{y,E}=(-1)^{\epsilon_y+1}$ in this case.
In either case, we can write
$$t_{y,E}=(-1)^{\epsilon_y+(w_2(E),C)}. \eqn\juxon$$

Now we can assemble the ingredients.  Pick a class $x\in H^2(M,{\bf Z}_2)$,
and let $\langle ~~~\rangle_{T,x}$
denote a topologically normalized correlation function, summed
over all isomorphism classes of $SO(3)$ bundles $E$ with $w_2(E)=x$.  Then
$$\eqalign{
&\left\langle\exp \left(\sum_a\alpha_aI(\Sigma_a)+\lambda \co\right)
\right\rangle_{T,x} \cr
=&~~~ 2^{1+{1\over 4}\left(7\chi + 11\sigma\right)}
\exp\left(\sum_{a,b}{\alpha_a\alpha_b\over 2}\#(\Sigma_a\cap \Sigma_b)
+2\lambda \right)\prod_{y}\left(e^{\phi_y}+(-1)^{(x,C_y)}
t_ye^{-\phi_y}\right)\cr
&~~~+i^{\Delta-x^2} 2^{1+{1\over 4}\left(7\chi + 11\sigma\right)}
\exp\left(-\sum_{a,b}{\alpha_a\alpha_b\over 2}\#(\Sigma_a\cap \Sigma_b)
-2\lambda \right)\prod_{y}\left(e^{-i\phi_y}+(-1)^{(x,C_y)}
t_ye^{i\phi_y}\right).
  \cr}         \eqn\onurkky$$
Thus, while the $SU(2)$ theory determines the cohomology classes
of the $C_y$ modulo torsion, the $SO(3)$ theory determines the
pairings of the $C_y$ with arbitrary mod two classes and thus
is sensitive to the two-torsion.  The additional information
is trivial for simply-connected $M$, but in general the $SO(3)$
theory does contain limited additional information.

That the right hand side of \onurkky\ is real follows from \rippo\ together
with the following standard fact:
$$(x,x)+(x,C)= 0~{\rm modulo} ~2,\eqn\curf$$
with $x$ any element of $H^2(M,{\bf Z}_2)$ and
$C=\sum_yC_y$ the canonical divisor (which we are identifying with
the Poincar\'e dual cohomology class).  Indeed, since $C$ reduces
mod 2 to the second Stieffel-Whitney class $w_2(M)$, this follows
from the Wu formula, which asserts that $(x,x)=(x,w_2(M))$ mod 2
for any $x\in H^2(M,{\bf Z}_2)$.  See [\kd, p. 6] for a quick proof
of the Wu formula for the case that $M$ is simply-connected.

\subsection{Groups Of Higher Rank}

Now, let us briefly discuss what happens for gauge groups of
higher rank.

One modification is that the global symmetry of the $N=2$
theory perturbed by a mass term that respects $N=1$ invariance
is not $\bz_4\times \bbz$ but $\bz_{2h}\times\bbz$, with $h$ the
dual Coxeter number.  In bulk this is broken down to
$\bz_2\times \bbz$, so the number of vacuum states in bulk  is
$h$ instead of 2.  Along the cosmic string it is most plausible
that $\bz_2\times \bbz$ is broken down to a diagonal $\bz_2$
subgroup (which permits fermion masses), giving a two-fold degeneracy.

A more substantial change comes from the fact that for the
gauge groups of rank greater than one, there are
higher Casimir operators.  For instance, for $G=SU(n)$,
\barnaby\ can be generalized to $\co^{(0)}_r(x)=
{1\over (2\pi)^r r!}\Tr\phi^{r+1}(x)$
with $1\leq r\leq n-1$.  Applying the descent procedure,
one gets operator valued $k$-forms $\co^{(k)}_r(x)$, for $r=1\dots n-1$.
The main novelty is the occurrence of non-trivial operator-valued
four-forms  (in contrast to $\co^{(4)}_1$ which is related to a classical
invariant, the instanton number).
One can perturb the Lagrangian by
$$L \to L+\sum_{r=2}^{n-1} b_r\int_M\co^{(4)}_r , \eqn\ghin$$
with arbitrary coupling constants $b_r$; this gives a family of
topological field theories.

We can think of the $b_r$ as
infinitesimal variables in the sense that $b_r{}^{n_r}=0$
for some large, unspecified values of $n_r$.
This is sufficient for studying the correlation functions, which
is all we really want.  Thus in particular, turning on the $b_r$
does not modify the mass gap and symmetry breaking
pattern, which are stable against small perturbations.
After turning on the $b_r$,
correlation functions of the $\co^{(d)}_r$ with $d<4$ can be
computed as we have done above for $SU(2)$ in terms of a certain
number of universal quantities  (in the $SU(2)$ case they
are $\eta,\langle\co\rangle,\langle V\rangle,a$, and $b$;
for higher rank there are more such constants because there are more
operators).  Once the $b_r$ are permitted to vary, the universal
quantities are no longer ``constants'' but
become instead functions of the $b_r$.\foot{One can see such
a phenomenon in the two-dimensional analog of Donaldson theory -- where
it occurs already for $SU(2)$.  The analog of the $b_r$ are
the $\delta_i$ in eqn. (5.5) of [\ugly]; the dependence on $\delta_i$
arises through a change of variables given in eqn. (5.9) as a result
of which the ``constants'' that control correlation functions of
$Q_{(0)}$ and $Q_{(1)}$ become functions of the $\delta_i$.}

Thus, a somewhat more elaborate structure arises in the case
of Lie groups of rank greater than one.  However, the implications
for topology of Kahler manifolds may be limited, since one can only
hope to detect the structure of $C$, but most of the
invariants of $C$ that possess the necessary properties
to be detected by theories such as these ones
are already detected by the $SU(2)$ and $SO(3)$ theories.
Perhaps with other groups one can learn more about the torsion
part of the cohomology classes of the $C_y$; so far we have
detected only the two-torsion (via the $SO(3)$ theory).

\chapter{Properties Of Supersymmetric Yang-Mills Theories}

The purpose of this section is to fill in details suppressed
in \S2 concerning the relevant
properties of supersymmetric Yang-Mills theories.  In doing so,
we will generally use conventions of Wess and Bagger [\wess].  For
instance, doublets of the $SU(2)_L$ (or $SU(2)_R$) rotation symmetries
are represented by spinor indices $\alpha,\beta\dots
=1,2$ (or $\dot\alpha,\dot\beta,\dots
=1,2$).  Doublets of the internal $SU(2)_I$ symmetry will be denoted
by indices $i,j,\dots=1,2$.  These indices are raised and lowered
with the antisymmetric tensor $\epsilon_{\alpha\beta}$ (or $\epsilon_{
\dot\alpha\dot\beta},\,\,\epsilon_{ij}$) with sign convention such that
$\epsilon_{12}=1=\epsilon^{21}$.
Tangent vector indices to spacetime are denoted $m,n,\dots =1\dots 4$;
the spinor and tangent vector indices are related by the tensor
$\sigma_{m\alpha\dot\alpha}$ described in appendix A of [\wess];
one similarly uses $\sigma_{mn\alpha}{}^\beta=\left(\sigma_{m\alpha
\dot\alpha}\sigma_n^{\beta\dot \alpha}-m\leftrightarrow n\right)/4$, etc.

The fields of the minimal $N=2$ supersymmetric Yang-Mills theory
are the following: a gauge field $A_m$, fermions $\lambda_{\alpha}{}^i$
and $\bar\lambda_{\dot\alpha i}$ transforming as $(1/2,0,1/2)$ and
$(0,1/2,1/2)$ under $SU(2)_L\times SU(2)_R\times SU(2)_I$, and
a complex scalar $B$ -- all in the adjoint representation of the gauge
group.  Covariant derivatives are defined by $D_m\Phi=(\partial_m+iA_m)\Phi$
and the Yang-Mills field strength is $F_{mn}=\partial_mA_n-\partial_nA_m
+[A_m,A_n]$.
The supersymmetry generators transform as $(1/2,0,1/2)\oplus
(0,1/2,1/2)$; introducing infinitesimal parameters $\eta_{\alpha}{}^i$
and $\bar\eta_{\dot\alpha i}$, they can be written
$$
\eqalign{
\delta A_m & =-i\bar\lambda^{\dot\alpha}{}_i\sigma_{m\alpha\dot\alpha}
\eta^{\alpha i}+i\bar\eta^{\dot\alpha}{}_i\sigma_{m\alpha\dot\alpha}
\lambda^{\alpha i} \cr
\delta \lambda_\alpha{}^i & =
\sigma^{mn}{}_{\alpha\beta}\eta^{\beta i} F_{mn}
+i\eta_\alpha{}^i D +i\sqrt 2\sigma^m{}_{\alpha\dot\alpha}D_mB\epsilon^{ij}
\bar\eta^{\dot\alpha}{}_j  \cr
\delta\bar\lambda_{\dot\alpha i} & = \sigma^{mn}{}_{\dot\alpha\dot\beta}
\bar\eta^{\dot\beta}{}_iF_{mn}-i\bar\eta_{\dot\alpha i} D +i\sqrt 2
D_m\bar B\epsilon_{ij}\eta^{\alpha j} \cr
\delta B & = \sqrt 2\eta^{\alpha i}\lambda_{\alpha i} \cr
\delta \bar B& = \sqrt 2
\bar\eta^{\dot\alpha}{}_i\bar\lambda_{\dot\alpha}{}^i
,\cr}      \eqn\transfor$$
with $D=[B,\bar B]$.  The minimal Lagrangian is (with signature $-+++$)
$$\eqalign{ L={1\over e^2}\int_Md^4x\Tr & \left({-1\over 4}F_{mn}F^{mn}
-i\bar\lambda^{\dot\alpha}{}_{i}\sigma^m_{\alpha\dot\alpha}D_m\lambda
^{\alpha i} -D_m\bar BD^m B\right.\cr
&\left.-{1\over 2}[\bar B,B]^2 -{i\over \sqrt 2}\bar B\epsilon_{ij}
[\lambda^{\alpha i},\lambda_\alpha{}^j]+{i\over \sqrt 2}B\epsilon^{ij}
[\bar\lambda_{\dot\alpha i},\bar\lambda^{\dot\alpha}{}_i]\right)
.\cr} \eqn\onorno$$
These formulas have been adapted from the formulas of [\wess, p. 50]
for $N=1$ supersymmetry.  Here $\Tr $ is an invariant quadratic form on the
Lie algebra which for $G=SU(N)$ we can conveniently take to be the
trace in the $N$ dimensional representation.

To construct a topological field theory, one replaces $SU(2)_R$
by a diagonal subgroup of $SU(2)_R\times
SU(2)_I$.  This can be conveniently
incorporated in the formalism by replacing the internal indices
$i,j,\dots$ by another $SU(2)_R$ index $\dot\beta$.
To describe the topological transformation laws, one sets
$$\eqalign{\eta^{\alpha i}&=0,\cr
\bar \eta^{\dot\alpha\dot\beta}&=-\epsilon^{\dot\alpha\dot\beta}\rho \cr}
\eqn\tuyt$$
(with $\rho$ an anticommuting parameter).  This gives a one-component
supersymmetry which is the BRST symmetry of the twisted topological
theory; the corresponding charge will be called $Q$.
The transformation laws are
$$\eqalign{\delta A_m & = i\rho \sigma_{m\alpha\dot\beta}\lambda^{\alpha
\dot\beta} \cr
\delta\lambda_{\alpha\dot\beta} & = i\rho\sqrt 2\sigma^m{}_{\alpha\dot\beta}
D_mB\cr
\delta\bar\lambda_{\dot\alpha\dot\beta}
& = \rho\sigma^{mn}{}_{\dot\alpha\dot
\beta}F_{mn}+i\epsilon_{\dot\alpha\dot\beta}\rho D \cr
\delta B & = 0 \cr
\delta \bar B & = -\sqrt 2 \rho\epsilon_{\dot\alpha\dot\beta}\bar\lambda
^{\dot\alpha\dot\beta} .\cr} \eqn\toptrans$$
The square of this transformation vanishes, up to a gauge transformation.
By requiring invariance under this topological symmetry, one can
construct a topological field theory on an arbitrary four-manifold $M$.

Now let us construct the observables of this theory.  If we set
$$\eqalign{
\lambda_m&=\sigma_m{}^{\alpha\dot\beta}\lambda_{\alpha\dot\beta}\cr
            B & = {i\phi\over 2\sqrt 2} , \cr}     \eqn\rurly$$
then the topological transformation laws become, in part,
$$\eqalign{  \delta A_m & = i\rho \lambda_m \cr
             \delta \lambda_m & = -\rho D_m\phi   \cr
             \delta\phi & = 0 ,\cr}     \eqn\mumsy$$
which is a standard topological field theory multiplet (related
mathematically to the equivariant cohomology of the gauge group
acting on the space of connections).  The other fields do not contribute
to the BRST cohomology.

The most obvious invariant operators
are of the form $\co(x)=P(\phi(x))$,
with $P$ an invariant polynomial on the Lie algebra of $G$ and $x$ a point
in the four-manifold $M$.  For $G=SU(2)$, the only essential choice
of $P$ is the quadratic Casimir operator, and then we can take
$$\co(x) = {1\over 8\pi^2}\Tr\phi^2=-{1\over \pi^2}\Tr B^2. \eqn\milmo$$
Calling this operator $\co^{(0)}(x)$ to stress that it can be regarded
as a zero-form on $M$, one next iteratively finds operator-valued
$k$-forms $\co^{(k)}$, $1\leq k \leq 4$, obeying
$$d\co^{(k)}=\{Q,\co^{(k+1)}\}.  \eqn\ifflox$$
For full details see [\witten].
In the case of a simply-connected four-manifold, the most important
of these operators is $\co^{(2)}$
(as we explained in \S2.2).  The components
of $\co^{(2)}$ were called $Z_{mn}$ in \S2; explicitly
$$Z_{mn}=-{i\over 4\pi^2}\Tr\left(2\sqrt 2 BF_{mn}+
\lambda_m\lambda_n   \right). \eqn\xoxo$$

One now, as in \S2, defines observables by
$$I(\Sigma)=\int_\Sigma \co^{(2)}, \eqn\fflox$$
with $\Sigma$ a two-dimensional homology cycle.  From \ifflox\ one deduces
that $I(\Sigma)$ is $Q$-invariant, and up to $\{Q,\dots\}$ depends only
on the homology class of $\Sigma$.  Equivalently, one can introduce
a closed two form $\theta$ and consider
$$I(\theta)=\int_M\co^{(2)}\wedge \theta. \eqn\glox$$
The two formulations are essentially equivalent, with $\theta$ being
the Poincar\'e dual of $\Sigma$.

\section{Kahler Manifolds}

Now we consider the further constructions that are possible if the metric
on $M$ is Kahler.  This means that the holonomy is not $SU(2)_L\times
SU(2)_R$ but $SU(2)_L\times U(1)_R$, with $U(1)_R$ a subgroup of
$SU(2)_R$.  The two-dimensional representation of $SU(2)_R$ decomposes
under $U(1)_R$ as a sum of two one-dimensional representations
(of ``charge $\pm 1/2$'').  Accordingly, there are invariant projections
of any doublet $v^{\dot\beta}$ onto the components $v^{\dot 1}$ and
$v^{\dot 2}$ transforming in definite $U(1)_R$ representations.
This permits one to define a complex structure on $M$; one simply
declares that the one-forms $dx^m\sigma_{m\alpha\dot 2}$ are of
type $(0,1)$ while the one-forms $dx^m\sigma_{m\alpha\dot 1}$ are of
type $(1,0)$.

Similarly, one can decompose the topological generator of equation
\tuyt\ into two components, taking
$$\eqalign{\eta_\alpha{}^i & = 0 \cr
       \bar\eta_{\dot\alpha \dot 1}&=\rho_1\epsilon_{\dot\alpha\dot 1} \cr
       \bar\eta_{\dot\alpha \dot 2}&=\rho_2\epsilon_{\dot\alpha\dot 2} \cr
       }\eqn\uyt$$
with anticommuting parameters $\rho_1,\rho_2$.  The explicit $\dot 1$
in \uyt\ means that the $\rho_1$ transformation is the part of the
topological symmetry of \tuyt\ that comes from an $N=1$ subalgebra
of the underlying $N=2$ supersymmetry; likewise the $\rho_2$
transformation comes from the second $N=1$ subalgebra.
Let $Q_1,$ $Q_2$ be the charges corresponding to the $\rho_1$ and
$\rho_2$
transformations.  Then $Q=Q_1+Q_2, $ and $Q_1{}^2=Q_2{}^2=\{Q_1,Q_2\}=
0$.

The vertex operators $\co^{(k)}$ are invariant under both the $\rho_1$
and $\rho_2$ transformations, modulo the equations of motion.  The
fact that the equations of motion enter here will be important presently.

In this paper, it will suffice
to consider the $\rho_1$ symmetry, with $\rho_2=0$.
The transformation laws are
$$\eqalign{
\delta A_m & =i\rho_1\sigma_{m\alpha \dot 1}\lambda^{\alpha\dot 1} \cr
\delta \lambda_{\alpha\dot 1} & =
i\sqrt 2\rho_1\sigma^m{}_{\alpha\dot 1} D_m B
              \cr
\delta\lambda_{\alpha\dot 2} & = 0 \cr
\delta\bar\lambda_{\dot\alpha\dot 1} & = \rho_1\left(\sigma^{mn}{}_{\dot
\alpha\dot 1}F_{mn}-i\epsilon_{\dot\alpha\dot 1} D\right) \cr
\delta\bar\lambda_{\dot\alpha\dot 2} & = 0 \cr
\delta B & = 0 \cr
\delta \bar B & = -\sqrt 2\rho_1\bar\lambda_{\dot 1\dot 2}.\cr}\eqn\guyt$$

Certain of these equations have the following interpretation.
If we decompose $F$ into its parts $F^{2,0}$, $F^{1,1}$, and $F^{0,2}$
of the indicated types, then $\delta\bar\lambda_{\dot 1\dot 1}$ is
proportional to $F^{0,2}$, so a BRST-invariant configuration must
have $F^{0,2}=0$; hence it describes a holomorphic vector bundle.
The first equation shows that the $(0,1)$ part of the connection
$A$ is BRST-invariant;
hence the holomorphic structure of the bundle is BRST-invariant.

To justify a claim
made at the end of \S2.4, the reason that $Q_1$-invariance
is enough in analyzing the topological correlation functions is that
the change in the Lagrangian under a change in the Kahler
metric or coupling constant can be written in the form
$\{Q_1,\dots\}$.  We can see this as follows.
Since $Q_1$ lies in an $N=1$ algebra,
it is convenient to recall the standard way of writing the $N=2$
theory in $N=1$ superspace.  The part involving the $\Phi$ multiplet
is $\int d^4x d^2\theta d^2\bar \theta
\Tr\bar\Phi \Phi$.  Since $Q_1$ acts (up to
a total derivative) as $\int d\theta^\alpha$ for suitable $\alpha$,
this is $\{Q_1,\dots\}$, and its variation with respect to metric
or coupling has the same property.  The gauge kinetic energy
is $\int d^4x d^2\theta\Tr W^2+\int d^4x\d^2\bar\theta\Tr\bar W^2$.
The first term here is $\{Q_1,\dots\}$ for the same reason as before,
while (by a standard identity)
the second coincides with the first up to a multiple of the
instanton number.  Hence a change in Kahler metric or gauge coupling
induces at most a term of the form $\{Q,\dots\}$ plus a change in
the $\theta$ angle (which is the coupling that multiplies the instanton
number).  Using the familiar chiral anomaly of the $U(1)$
$R$-current, a change in the $\theta$ angle can be absorbed in a chiral
rotation or rescaling of the fields $I(\Sigma)$ and $\co$
and (allowing for the gravitational
part of the chiral anomaly) a redefinition of the constants called
$a$ and $b$ in \uddu.
The rescaling of $I(\Sigma)$ and $\co$ merely affect the values
of the constants $\eta,\langle\co\rangle$, and $v$ that entered our
formulas.

\subsection{The Mass Term}

Exploiting the fact that $Q_1$ is the only essential symmetry,
we want to introduce a mass term for some of the fields, preserving
the $Q_1$ symmetry.  As was foreseen in \S2, this construction depends
on a choice of holomorphic two-form $\omega$ on $M$.
We consider adding to the Lagrangian $I(\omega)=\int_M\co^{(2)}\wedge
\omega$.  This contains a term $\int_M\Tr BF\wedge \omega$. Since
$\omega$ is of type $(2,0)$, only the $(0,2)$ part of $F$ enters here;
as noted after equation \guyt, this is of the form $\{Q_1,\dots\}$.
So modulo $\{Q_1,\dots\}$, the $BF$ term can be dropped and
$I(\omega)$ is proportional to
$$L_1=-{1/2}\int_M\Tr\lambda_\alpha{}^{\dot 2}\lambda^{\alpha\dot 2}
\sigma_{mn \dot 2\dot 2}\cdot \omega_{kl}\epsilon^{mnkl}\sqrt g d^4x.
    \eqn\jurry$$

Now, in verifying the $Q_1$ invariance of $L_1$ (or equivalently,
of $I(\omega)$) it is necessary to use the equations of motion.  In
fact, a small computation shows that
$$\delta L_1=-i\sqrt 2 \rho_1\int_M\Tr B\epsilon^{\dot\alpha\dot 2}
\sigma^m{}_{\alpha\dot\alpha}D_m\lambda^{\alpha\dot 2}\sigma_{np\dot 2
\dot 2}\omega_{kl}\epsilon^{npkl}\sqrt g d^4x.  \eqn\jippo$$
This vanishes by the equation of motion of $\bar\lambda$, as one
can see by considering the part of the original Lagrangian $L$ that
contains $\bar\lambda$.  This is
$$L_0=-{i\over e^2}
\int d^4x\sqrt g\Tr \left(\bar\lambda^{\dot\alpha}_{\dot 2}
\sigma^m{}_{\alpha\dot\alpha}D_m\lambda^{\alpha\dot 2}+\sqrt 2
\bar\lambda_{\alpha\dot 2}[B,\bar \lambda^{\dot\alpha\dot 2}]\right).
\eqn\urnamy$$
Instead of saying that $\delta L_1$ is $Q_1$ invariant modulo the
equations of motion derived from $L$, one can modify the $Q_1$
transformation laws so that the full Lagrangian $L+L_1$
is $Q_1$-invariant.  The requisite correction to the transformation
laws is that, instead of vanishing, $\delta \bar\lambda_{\dot\alpha\dot 2}$
should be as follows:
$$\delta\bar\lambda^{\dot\alpha}{}_{\dot 2}=-\sqrt 2 e^2\rho_1B
\epsilon^{\dot \alpha \dot 2}
\sigma_{np\dot 2\dot 2}\omega_{kl}\epsilon^{npkl}
.\eqn\cortrans$$
Since $\delta B=0$,
this correction to the transformation law does not affect the
fact that $Q_1{}^2=0$.  Nor does it affect the
$Q_1$ invariance of the various observables of the theory
(such as $\co$ or $I(\Sigma)$), as they are independent of $\bar\lambda$.

While preserving $Q_1$ invariance, we could  add to the
Lagrangian any expression of the form $\{Q_1,V\}$.
It is convenient to do so with the particular choice
$$V=-{1\over \sqrt 2}\int d^4x\sqrt g\Tr\bar B\,
\bar\lambda^{\dot\alpha}{}_{\dot 2}\epsilon^{\dot\beta\dot 2}
\sigma_{np\dot\alpha\dot\beta}\bar\omega_{kl}\epsilon^{npkl}.\eqn\hcoxn$$
The corresponding addition to the Lagrangian is
$$L_2=-{1\over 2}\int d^4x\sqrt g\bar\lambda^{\dot\sigma}{}_{\dot 2}
\bar\lambda_{\dot\sigma\dot 2}\sigma_{np}{}^{\dot 2\dot 2}\bar
\omega_{kl}\epsilon^{npkl}-e^2\int d^4x\sqrt g\Tr \bar B B
\cdot\left|\sigma_{np\dot 2\dot 2}\omega_{kl}\epsilon^{npkl}\right|^2
. \eqn\huff$$

So we consider the combined Lagrangian
$$\eqalign{\widehat L=&L+L_1+L_2  =L+I(\omega)+\{Q_1,\dots\} \cr
=& L-{1\over 2}\int_Md^4x\Tr\left(m\lambda_\alpha{}^{\dot 2}\lambda^\alpha{}
^{\dot 2}+\bar m \bar\lambda^{\dot\sigma}{}_{\dot 2}\bar\lambda_{\dot
\sigma\dot 2}\right)-e^2\int_Md^4x\sqrt 2\bar m m \Tr \bar B B
.\cr}       \eqn\uff$$
with
$$m = \sigma_{mn\dot 2\dot 2}\omega_{kl}\epsilon^{mnkl}.\eqn\curry$$
As promised in \S2, we have realized a mass term for the $N=
1$ matter multiplet (which
consists of $B$ and $\psi_\alpha=
\lambda_\alpha{}^{\dot 2}$) by adding to the Lagrangian a term of the
form $I(\omega)+\{Q_1,\dots\}$.  The mass is proportional to the
holomorphic two-form $\omega$, as expected.  The $N=1$ gauge
multiplet, consisting of the gauge field $A_m$ and the gluino
$\lambda_\alpha=\lambda_\alpha{}^{\dot 1}$, remains massless.

\subsection{Symmetries}

For subsequent use, it will be important to determine the
transformation law of the observables $\co$ and $I(\theta)$
under the global symmetries of the theory.  (For a related
analysis, see [\park].)

As we discussed in \S2, the twisted theory, perturbed by the mass
term as in \uff, has a global symmetry $\bz_4\times \bbz$.
We will describe the symmetries in the language of $N=1$ supersymmetry,
the $N=1$ multiplets being $(A_m,\lambda)$ and $(B,\psi)$.
The generator $\alpha$ of $\bz_4$ acts by
$$\alpha(\lambda)=i\lambda,~~\alpha(\bar\lambda)=-i\bar\lambda,
{}~~\alpha(B)=iB, \eqn\nuggo$$
with $A_m,\psi$ being invariant.  (Such a symmetry, which
acts differently on different components of a supermultiplet,
is usually called an $R$ symmetry.)  The non-trivial element $\beta$
of $\bbz$ acts by
$$\beta(B)=-B,~~~\beta(\psi)=-\psi,\eqn\jukko$$
with the gauge multiplet $(A_m,\lambda)$ being invariant.

Now let us determine the transformation laws of the observables.
$\co(x)=-\Tr B^2/\pi^2$ evidently transforms as $(-1,1)$ under $\alpha$
and $\beta.$
Now let us consider the quantum numbers of $I(\theta)$ with
$\theta$ a $(p,q)$-form (with $p+q=2$); we will schematically call
this operator $I^{p,q}$.
$I^{1,1}$ is of the general form $\int_M\theta\wedge (\lambda
\psi +BF)$; therefore, from the transformation laws presented
in the last paragraph, $I^{1,1}$ transforms as $(i,-1)$.

What about $I^{2,0}$?  This is analyzed most efficiently by noting
that $I^{2,0}$ is present in the Lagrangian, so it must be invariant,
transforming
as $(1,1)$, modulo $\{Q_1,\dots\}$.  If one tries to verify this
explicitly,  one has schematically $I^{2,0}\cong \psi\psi+BF$ in the original
twisted topological field theory.
The two terms transform differently under the symmetries, but the
second is $\{Q_1,\dots\}$, a fact exploited above, and so can be dropped.
It appears
that we can take $I^{2,0}\cong\psi\psi$, which is invariant,
as expected.  This is the correct structure of $I^{2,0}$ in the absence
of the mass perturbation introduced above, but that perturbation brings
about the following change.  We recall that $I^{2,0}$ was only
$Q_1$-invariant modulo the $\bar\lambda$ equations of motion.  In
adding a perturbation $L_2=\{Q_1,V\}$ to the Lagrangian to complete
the mass term \uff, we have disturbed the $\bar\lambda$ equations of
motion, and accordingly an extra term must be added to $I^{2,0}$.
The result can be found most simply by noting that (before or after
addition of $L_2$) $I^{2,0}$ can be interpreted as $\partial\widehat L/
\partial m$, with $m$ as in \curry.  This gives an extra term
$\partial L_2/\partial m$, and finally $I^{2,0}\cong \psi\psi+
\bar\omega\bar B B$.

Finally, we consider
$I^{0,2}$, which is of the general form $\int_M\theta\wedge(\lambda\lambda
+BF)$.  Again, the two terms transform differently
under the symmetries.  The second term can be written as
$\{Q_1,\dots\}$ using the correction \cortrans\ to the transformation
laws,\foot{This is actually only true for $\theta=\omega$, but that
is good enough for probing Donaldson theory as $\omega$ was in any
case an arbitrary holomorphic two-form.}
and so is again inessential.  So $I^{0,2}$ transforms
as $(-1,1)$.

To summarize this discussion, the quantum numbers of the observables
are as follows:
$$\eqalign{\co: & ~~~(-1,1)  \cr
           I^{2,0}: & ~~~ (1,1) \cr
           I^{1,1}: & ~~~ (i,-1) \cr
           I^{0,2}: & ~~~ (-1,1). \cr }  \eqn\poxy$$
Moreover modulo $\{Q_1,\dots\}$, we have schematically
$$\eqalign{
            I^{1,1}&\cong \lambda\psi+BF \cr
            I^{0,2} & \cong \lambda\lambda  . \cr}   \eqn\oxy$$

\subsection{Gluino Condensation}

Let us now analyze in bulk the dynamical effects of the mass perturbation
in \uff.  This means that we work on a flat ${\bf R}^4$ and
take $\omega=dz_1\wedge dz_2$; of course any example, suitably
scaled up, looks like this locally as long as one keeps away from the
zeroes of $\omega$.

Seen as an $N=1$ theory, the $N=2$ supersymmetric Yang-Mills
theory has a gauge multiplet $(A_m,\lambda)$ and a matter
multiplet $\Phi=(B,\psi)$.  The perturbation in \uff\ is in bulk
simply the usual $N=1$ invariant bare mass term for the $\Phi$ multiplet.
Its addition to the Lagrangian leaves at low energies the pure
$N=1$ supersymmetric gauge theory of $(A_m,\lambda)$.  This
theory has a $\bz_4$ symmetry, generated by the transformation $\alpha$
described in \nuggo.  It is believed that this symmetry is broken
to its $\bz_2$ subgroup, by ``gluino condensation,'' that is,
by the expectation value of $\Tr\lambda^\alpha
\lambda_\alpha$, which is essentially our friend $I^{0,2}$.
We will abbreviate $\Tr\lambda^\alpha\lambda_\alpha$ as $\lambda\lambda$.

It will be necessary therefore for us to discuss some standard properties
of gluino condensation.  Let us recall the basic structure of the
renormalization group.  If a quantum theory is defined using
a renormalization point $\mu$ and with a coupling constant $g$,
then the effective coupling constant $\bar g$ obeys
$$\left(\mu{\partial\over\partial\mu}+\beta(g){\partial\over\partial g}
\right)\bar g = 0 ,\eqn\juyr$$
with $\beta(g)$ the beta function.  If $\bar g$ is defined by measurements
at an energy $E$, then $\bar g$ is a function only of the dimensionless
variables $E/\mu$ and $g$.  If for weak coupling the beta
function looks like $\beta=-bg^3+\dots$ (with $b>0$ for asymptotically
free theories) then the renormalization group equation can be written
in the weak coupling regime as
$$\left(\mu{\partial\over\partial\mu}-2b(g^2)^2{\partial\over\partial
g^2}\right)\bar g^2(E/\mu,g) = 0.  \eqn\hucc$$
The solution is
$${1\over \bar g^2(E/\mu)}={1\over g^2}-2b\ln (\mu/E).  \eqn\oppo$$
Strong coupling arises for $\bar g\cong 1$ or
$${E}\cong \mu\exp(-1/2bg^2). \eqn\noppo$$
Even though a weak coupling approximation was used in solving \juyr,
\noppo\ gives correctly the singular behavior of $E$ for small $g$
which is in fact dominated by the behavior in the weakly coupled regime.

Now, we are interested in the case in which at some mass scale
$m$, an $N=2$ theory is explicitly broken down to an $N=1$ theory.
The $N=2$ theory has a beta function $\beta=-bg^3+\dots$,
and the $N=1$ theory has a beta function $\beta=-b'g^3+\dots$.
The relation between them is
$$b={2b'\over 3}.\eqn\uppu$$
(In fact, the beta function of supersymmetric Yang-Mills theory of
general $N$ is proportional to $N-4$ and has a well-known zero for
$N=4$.)
We suppose that $m<<\mu$ and is large enough to be in the weak coupling
regime.  In that case, as long as $E\geq m$,
we can use the solution \oppo\ for the effective coupling.
However, for $E\leq m$, we must evolve the effective coupling using
the beta function of the low energy theory.
In doing this, we use \oppo\ to determine the initial conditions
for the renormalization group evolution of the low energy theory.
In fact, for the renormalization scale $\mu_L$ and coupling $g_L$
of the low energy theory, we take the values
$\mu_L=m$ and $g_L=\bar g(m/\mu,g)$
that are appropriate for the microscopic $N=2 $ theory at the scale
$m$.  The subsequent evolution of the effective coupling to energies
below $m$ is carried out using
an equation just like \juyr, with $\mu,g,\bar g$, and $\beta$ replaced by $m,
\bar g(m/\mu,g)$, the effective coupling $\bar g_L$ of the low
energy theory, and the beta function of the low energy theory.
So we get the analog of \oppo:
$${1\over \bar g_L{}^2(E/m,\bar g^2(m/\mu,g))}={1\over \bar g^2(m/\mu,g)}
-2b'\ln(m/E). \eqn\noppo$$
These equations combine to give
$${1\over \bar g_L{}^2}={1\over g^2}-2b\ln (\mu/m)
-2b'\ln(m/E).     \eqn\koppo$$
Gluino condensation occurs for $\bar g_L$ of order one, which occurs
for
$$E\cong \exp(-1/2b'g^2)m^{1-(b/b')}\mu^{b/b'}. \eqn\doppo$$

\REF\newseiberg{N.  Seiberg, ``Naturalness Versus
Supersymmetric Nonrenormalization Theorems,''
Phys. Lett. {\bf 318B} (1993) 469.}
In our particular case, using
\uppu, the dependence of the energy scale on $m$ and $\mu$
is
$$E\cong m^{1/3}\mu^{2/3}.\eqn\froppo$$
Since the gluino condensate has dimension three and
so scales as $E^3$, we get
$$\langle \lambda\lambda\rangle \cong m \mu^2, \eqn\occro$$
a result that will be essential later.
This estimate of the $m$ dependence of $\langle\lambda\lambda\rangle$
is reliable for $m$ in the weak coupling regime -- a harmless restriction
in the topological theory as the coupling constant dependence of the
Lagrangian is $\{Q,\dots\}$.
It is possible to show that the $m$ dependence given in \occro\
is exact using arguments of holomorphicity, as in [\cern,\newseiberg].

Of course, in \occro\ we have determined only the $m$ dependence
of the gluino condensate, not the overall constant multiplying this.
Because of the basic structure of $\bz_4$ broken to $\bz_2$, there
are really two vacua with $\langle\lambda\lambda\rangle\cong
\pm m\mu^2$, times a constant independent of $m$ and $\mu$.

\section{Evaluation Of Correlation Functions In Bulk}

Apart from gluino condensation, the $N=1$ theory is also
believed to have a mass gap.  In \S2.3, we sketched
how the topological correlation functions are determined by
the mass gap.  I will here fill in various details omitted
there, involving the quantum numbers of the operators, the precise
role of gluino condensation, and the lack of Lorentz invariance
(even in bulk) of the mass term that breaks $N=2$ down to $N=1$.
The reason for the latter statement is that, in the twisted theory,
the mass term comes from a choice of holomorphic two-form
(say $\omega=dz_1\wedge dz_2$ on a flat ${\bf R}^4$) which breaks
the $SO(4)$ rotation symmetry of ${\bf R}^4$ down to $SU(2)$.

\FIG\feydi{A one loop diagram by which the operator $\co$ can
be expressed as an effective operator in the low energy theory.}
First we consider correlation functions of the operator $\co=-\Tr B^2/
\pi^2$.
This operator transforms as $(-1,1)$, so it can get an expectation
value once the global symmetry $\bz_4\times\bbz$ is spontaneously
broken to $\bz_2\times\bbz$; this expectation value will, of course,
be odd under the broken symmetry.  To see how this comes about
more explicitly, we integrate out the massive $(B,\psi)$ multiplet
to express $\co$ purely in terms of light fields.  This can be done
via the one loop Feynman diagram of figure (\feydi), with the
result that $\co$ can be replaced by $\lambda\lambda/m$,
times a constant that I will not compute here.\foot{The factor of $1/m$
is needed on dimensional grounds: $\co$ and $\lambda
\lambda$ have dimension 2 and 3, respectively.
By examining the diagram, it is easy to see that one gets $1/m$
and not $1/m^{1-a}\bar m^a$ for $a\not= 0$; this actually follows
from general principles of holomorphicity.}  On the other hand, according
to \occro, the expectation value of $\lambda\lambda$
scales as $m\mu^2$.  So we get
$$\langle\co\rangle= \pm \mu^2 C, \eqn\ponno$$
with the $\pm$ sign depending on the sign of the gluino condensate
and $C$ a universal constant; $\mu^2C$ is the universal constant
$\langle\co\rangle$ of \S2.

\subsection{Two-Forms}

As a prelude to incorporating the two-form operators, let us compare
correlation functions $\langle~~~\rangle$ of topological observables
in the twisted theory to correlation functions $\langle ~~~\rangle_1$
in the theory perturbed by a mass term.  The relation is
$$\langle A_1A_2\dots A_n\rangle_1 =\langle A_1A_2\dots A_n e^{I(\omega)}
\rangle, \eqn\udd$$
since the mass term was $-I(\omega)+\{Q_1\dots\}$.

The main difference between the present problem and the case considered
in \S2.3 is the lack of Lorentz invariance in the bulk theory,
once a mass term is introduced via a choice of holomorphic two-form $\omega$.
In \S2.3, Lorentz invariance was used to prove that the vacuum
expectation value of $I(\Sigma)$ or equivalently $I(\theta)$ vanishes.
This will no longer be true.

If we look at the quantum numbers in \poxy, and take into account
the fact that $\bz_4\times \bbz$ is spontaneously broken down
to $\bz_2\times \bbz$, it is clear that $I^{1,1}$ cannot have
a vacuum expectation value, and it appears that either $I^{0,2}$
or $I^{2,0}$ might have expectation values.

In fact, the unbroken supersymmetry of the $N=1$ theory
implies that $\langle 1\rangle_1=0$ in bulk and hence that
$$\langle\exp I(\omega)\rangle = 0 \eqn\iccoc$$
for $\omega$ a holomorphic two-form.  Thus, $I^{2,0}$ has
no vacuum expectation value.

However, according to \oxy, $I^{0,2}$ is proportional to our
friend the gluino bilinear $\lambda\lambda$
and thus has a vacuum expectation value due to gluino condensation.
According to \occro, the expectation value of
$\lambda\lambda$ is proportional to the mass $m$ or in other
words to the two-form $\omega$.  The operator $I^{0,2}(\theta) =
\int_M Z\wedge \theta$ is of course also proportional to the two-form
$\theta$.  So one gets in fact
$$\langle I(\theta)\rangle_1=\eta_0\int_M\theta\wedge\omega, \eqn\ipsso$$
with $\eta_0$ a universal constant that is proportional to the
gluino condensate and therefore odd under the spontaneously broken
symmetry.  This is equivalent to
$$\langle I(\theta)e^{I(\omega)}\rangle = \eta_0\int_M\theta\wedge\omega.
 \eqn\nipsso$$
A consequence of \ipsso\ is as follows. If $\Sigma$ is any
Riemann surface in $M$ (not necessarily holomorphically embedded) then
$$ \langle I(\Sigma)\rangle_1=\eta_0\int_\Sigma\omega. \eqn\pixxi$$
This arises upon expanding the Poincar\'e dual of $\Sigma$ in classes
of type $(2,0)$, $(1,1)$, and $(0,2)$, recalling that $I^{p,q}$
has an expectation value only for $(p,q)=(0,2)$, and using
\ipsso.

Now let us consider the general case of several two-forms.
We still, however, temporarily keep away from zeros of the holomorphic
two-form $\omega$; those will be considered later.

As in \S2.3, to exploit the mass gap it is convenient to
introduce $s$ Riemann surfaces $\Sigma_1,\dots,\Sigma_s$ in space-time,
which we can assume to intersect only pairwise.  (The $\Sigma_a$
are not necessarily holomorphically embedded.)
As we saw in \S2.3, using Lorentz invariance of the bulk theory
and assuming that the mass gap holds everywhere, one gets
$$\left\langle
\exp\left(\sum_a\alpha_aI(\Sigma_a)\right)\right\rangle_1
 = \exp\left(\eta\sum_{a,b}
\alpha_a\alpha_b   \#I(\Sigma_a\cap\Sigma_b)\right).\eqn\burff$$
with some universal constant $\eta$.
We recall that in the derivation of \burff, one uses the fact that
(by Lorentz invariance) the expectation value of $I(\Sigma_a)$ vanishes
in bulk.  In the present case, that is no longer so;
the expectation values are given in \pixxi.  Including
the appropriate extra contributions, we get
$$\left\langle
\exp\left(\sum_a\alpha_aI(\Sigma_a)\right)\right\rangle_1
 = \exp\left(\eta\sum_{a,b}
\alpha_a\alpha_b   \#I(\Sigma_a\cap\Sigma_b)
+\eta_0\sum_a\alpha_a\int_{\Sigma_a}\omega\right),\eqn\bburff$$
just as if $I(\Sigma)$ is a free or Gaussian field with a one point
function given in \pixxi\ and a two point fuction given in \ombbo.

This is equivalent to
$$\left\langle
\exp\left(\sum_a\alpha_aI(\Sigma_a)+I(\omega)\right)\right\rangle
 = \exp\left(\eta\sum_{a,b}
\alpha_a\alpha_b   \#I(\Sigma_a\cap\Sigma_b)
+\eta_0\sum_a\alpha_a\int_{\Sigma_a}\omega\right).\eqn\bburff$$
It must be the case that $\eta_0=\eta$, since otherwise the formula
is not invariant under a change in the unnatural splitting of
the exponent on the left hand side
between $I(\omega)$ and the $I(\Sigma_a)$.  For
$\eta_0=\eta$, we can reexpress \bburff\ in a more invariant form
by expanding $I(\omega)$ in terms
of the $I(\Sigma_a)$ (plus $\{Q_1,\dots\}$):
$$\left\langle
\exp\left(\sum_a\alpha_aI(\Sigma_a)\right)\right\rangle
 = \exp\left(\eta\sum_{a,b}
\alpha_a\alpha_b  \#I(\Sigma_a\cap\Sigma_b) \right).\eqn\bburff$$
Thus, we have recovered the key formula \cvor\
of \S2.3, though in a more roundabout way
because of the lack of manifest Lorentz invariance
in the $N=1$ formulation.

Before leaving this subject, I will comment briefly on the
computation of $\eta$.  Near intersections of $\Sigma$'s, one
can assume that locally the $\Sigma$'s look like holomorphically
embedded Riemann
surfaces.  For instance, locally $\Sigma_1$ could be the locus $z_1=0$
and $\Sigma_2$ the locus $z_2=0$ in ${\bf R}^4\cong {\bf C}^2$.
In this case, $I(\Sigma_1)=\int_{z_1=0}I^{1,1}$ and similarly
for $I(\Sigma_2)$.  Since $I^{1,1}\cong \Tr (\lambda\psi+BF)$,
we have to evaluate something like
$$\int_{\Sigma_1\times\Sigma_2}\Tr(\lambda\psi+BF)(z_1)
\cdot \Tr(\lambda\psi+BF)(z_2).       \eqn\bicom$$
The role of the mass gap is here very explicit since $B,\psi$
have bare masses in the $N=1$ theory.  Integrating out these
massive fields, the integral in \bicom\ can be expressed in terms
of operators of the low energy theory -- that is, the $N=1$
theory of the gauge multiplet $A_m,\lambda$.  The leading contribution
comes from integrating out the $\psi$ field in \bicom\ and
is simply the local operator $\lambda\lambda/m$ evaluated at $z_1=z_2=0$.
Since the gluino condensate is
of the general form $\eta m$ with $\eta$ a constant independent of $m$,
the factors of $m$ cancel out and the local contribution to the
expectation value of \bicom\ is independent of $m$, as expected.
Of course, this contribution is odd under the broken symmetry,
as promised in \S2; the underlying reason for this is that
the operator $I^{1,1}$ transforms as $(i,-1)$ according to
\poxy.

\section{Behavior Near Cosmic Strings}

Now we want to analyze the behavior near a cosmic string.
We work on flat ${\bf R}^4$ with Euclidean coordinates
$y^1\dots y^4$ and complex coordinates $z_1=y^1+iy^2$, $z_2=y^3+iy^4$.
With the conventions of Wess and Bagger [\wess] (rotated
from Minkowski space to Euclidean space by setting $y^0=-iy^4$),
the Dirac operator
is
$$\sigma^mD_m=2\left(\matrix{{D\over Dz_2} & {D\over Dz_1} \cr
                     {D\over D\bar z_1} & -{D\over D\bar z_2}\cr}\right).
                         \eqn\urry$$
We consider the cosmic string defined by the holomorphic two-form
$\omega=z_1dz_1\wedge dz_2$, which vanishes on the curve $C$ defined
by $z_1=0$.  Away from $C$, the field $\psi^{\alpha}=\lambda^{\alpha
\dot 2}$ has a bare mass.  This mass vanishes in the core of the string.

As we claimed in \S2.7, there is a normalizable zero mode for
the $\psi$ field in the $z_1$ plane.   In fact, the equations
of motion of $\psi^1$ and $\bar \psi^{\dot 1}$, allowing for the
kinetic energy and the coupling to $\omega$, are
$$\eqalign{-i{\partial\psi^1\over\partial z_1}-\bar z_1\bar\psi^{\dot 1}
 & = 0 \cr
  -i{\partial \bar\psi^{\dot 1}\over\partial \bar z_1} +z_1\psi^1 & = 0.\cr}
              \eqn\guffrid$$
The zero mode is $\psi^1=\exp(-|z_1|^2)$, $\bar\psi^{\dot 1}=
i\exp(-|z_1|^2)$, with $\psi^2=\bar\psi^{\dot 2}=0$.

Now we wish to consider
the motion in the $z_2$ direction.  To do so,  we write
$\psi^1=\exp(-|z_1|^2)\psi'(z_2,\bar z_2)$, $\bar\psi^{\dot 1}=
i\exp(-|z_1|^2)\psi'(z_2,\bar z_2)$, with $\psi'$ a two-dimensional
fermi field, defined on $C$, and with values in the adjoint representation.
With this ansatz, the kinetic energy of the $\psi$ field reduces to
$$\int_C dz_2d\bar z_2\,\,\,
\Tr \psi'{D\over Dz_2}\psi'. \eqn\kiopd$$
This shows that, as claimed in \S2.7, $\psi'$ behaves like
a chiral fermi field along $C$.

To determine the spin structure to which $\psi'$ is coupled,
we consider a global situation
in which $C$ is a Riemann surface in a four-manifold $M$ defined
by the vanishing of a holomorphic two-form $\omega$.
Locally one can write $\omega = z_1dz_1\wedge dz_2$, with $z_1$
and $z_2$ being local coordinates normal and tangent to $C$.
The existence of a global two-form of the stated behavior implies
that $dz_1$ transforms as $(dz_2)^{-1/2}$.  To say this in a more
sophisticated way, the normal bundle $N_C$ to $C$ obeys
$N_C{}^{\otimes 2}\cong K_C{}^{-1}$, with $K_C$ the canonical
bundle to $C$.  Thus a section of $N_C$ can be interpreted as
a $(-1/2,0)$ form on $C$.  Hence $N_C{}^{-1}$ -- whose sections
transform as $\partial/\partial z_1$ -- is a spin bundle of $C$,
that is, a bundle of $(1/2,0)$ forms.  The complex conjugate
bundle $\bar N_C{}^{-1}$ is a bundle of $(0,1/2)$ forms or an
antiholomorphic spin bundle.

On the other hand, the formula \guyt\ shows that $\psi'$ transforms
as $DB/D\bar z_1$, to which it is related by the symmetry.
Thus, $\psi'$ transforms as a section of the antiholomorphic
spin bundle $\bar N_C{}^{-1}$, as was claimed in \S2.8.

\subsection{Quantum Numbers Of $V$}

The last claim made about cosmic strings in \S2
was that the operator $V$ has the quantum numbers claimed in \nurdox.

In fact, as in equation \jury, $V$ is the
contribution to integrals such as $I(\theta)
=\int_M\theta \wedge Z$ coming from the intersection with a cosmic
string.  As the Poincar\'e dual of the cosmic string is a class of
type $(1,1)$, only the $(1,1)$ part of $\theta$ contributes;
hence $V$ transforms under the symmetry as does $I^{1,1}$,
and \nurdox\ is a consequence of \poxy.

\section{A Useful Formula}

Finally, we must establish the formula \rippo\ that was needed to show
that the final expression for the Donaldson invariants is real.

We let $\omega$ be a holomorphic two-form on $M$ that vanishes
on a divisor $C$.  Let $K$ be the canonical bundle of $M$ and $K|_C$
its restriction to $C$. There is an exact sequence of sheaves
$$0\to \co\underarrow{\omega} K\to K|_C\to 0.\eqn\hunky$$
The first map is $f\to f\cdot \omega$.
This leads to an exact sequence
$$0\to H^0(M,\co)\underarrow{\omega} H^0(K)\to H^0(C,K|_C)
\to H^1(M,\co)\underarrow{\omega^1} H^1(M,K)\to \dots \eqn\hungry$$
We truncate this to
$$0\to H^0(M,\co)\underarrow{\omega} H^0(K)\to H^0(C,K|_C)
\to \ker(\omega^1) \to 0.      \eqn\hungry$$
The existence of such an exact sequence implies that
$$\dim H^0(M,\co)-\dim H^0(M,K)+\dim H^0(C,K|_C)-\dim \ker(\omega^1)=0.
\eqn\heffel$$

The map $\omega^1$
can be interpreted as the map from
$\alpha\in H^1(M,\co)\cong H^{0,1}(M,{\bf C})$
to $\alpha\wedge\omega\in H^1(M,K)\cong H^{2,1}(M,{\bf C})$.
A natural antisymmetric bilinear form on $H^{0,1}(M)$ can be defined by
$$\langle \alpha,\beta\rangle=\int_M\alpha\wedge\beta\wedge \omega.
\eqn\upply$$
By Poincar\'e duality,
$\alpha\wedge\omega$ vanishes as an element of $H^{2,1}(M,{\bf C})$
if and only if $\langle\alpha,\beta\rangle=0$ for all $\beta$.

Now in general, if $V$ is a vector space of finite dimension and
$\langle~~,~~\rangle$ is an antisymmetric bilinear form on $V$,
then the kernel of $\langle~~,~~\rangle$ is defined as the set of $x\in V$
such that $\langle x,y\rangle = 0$ for all $y\in V$; simple linear
algebra shows that the dimension of the kernel is congruent to the
dimension of $V$ modulo two.  The facts stated in the last paragraph
mean that $\ker \omega^1$ is the same as the kernel of $\langle~~,~~\rangle$
and hence that
$$\dim \ker \omega^1\cong \dim H^{0,1}(M)~~{\rm mod}~2. \eqn\mimblo$$
This lets us rewrite \heffel\ in the form
$$\dim H^0(C,K|_C)\cong 1-h^{0,1}(M)+h^{0,2}(M)~{\rm mod}~2.\eqn\piblo$$
(We recall that $h^{p,q}(M)={\rm dim}H^{p,q}(M)$.)

On the other hand, if $C$ is the union of connected components
$C_y$, then $H^0(C,K|_C)=\oplus_y H^0(C_y,K|_{C_y})$.
Above we showed in essence that $K|_{C_y}$ is the spin bundle of
the fermions that propagate along $C_y$ (actually, the chiral
fermions $\psi'$ are coupled to the antiholomorphic spin
bundle that is complex conjugate to $K|_{C_y}$).
So the number of fermion zero modes along $C_y$ is
$\epsilon_y=\dim H^0(C_y,K|_{C_y})$.
Taking account of the formula for $\Delta$ in \igaro, it follows
then that \piblo\ is equivalent to
$$\Delta \cong \sum_y\epsilon_y~{\rm mod}~2. \eqn\nibblo$$
This is the desired formula used in \S2.

\figout
\endpage
\refout
\end